# The Autohumidification Polymer Electrolyte Membrane Fuel Cell

**Jay B. Benziger, J. Moxley, S. Tulyani, A. Turner, A.B. Bocarsly, Y.G. Kevrekidis**
Princeton University
Princeton, NJ 08544

## Abstract

A PEM fuel cell was specially constructed to determine kinetics under conditions of well-defined gas phase composition and cell temperature. Steady state multiplicity was discovered in the autohumidification PEM fuel cell, resulting from a balance between water production and water removal. Ignition was observed in the PEM fuel cell for a critical water activity of ~0.1. Ignition is a consequence of the exponential increase of proton conductivity with water activity, which creates an autocatalytic feedback between the water production and the proton conduction. The steady state current in the ignited state decreases with increasing temperature between 50-105°C. At temperatures of >70°C five steady states were observed in the PEM fuel cell. The steady state performance has been followed with variable load resistance and hysteresis loops have been mapped. The dynamics of transitions between steady states are slow ~$10^3 - 10^4$ s. These slow dynamics are suggested to result from a coupling of mechanical and chemical properties of the membrane electrode assembly due to swelling of the membrane with water absorption.

## I. INTRODUCTION

Polymer electrolyte membrane (PEM) fuel cells are hydrogen fuel cells, where protons are conducted through an ionic polymer (ionomers). The original PEM fuel cells were based on polystyrene sulfonate. Since the early 1990's PEM fuel cells have employed fluorinated polymers functionalized with perfluorosulfonic acid groups such as DuPont's Nafion™. The basic function of the fuel cell is illustrated in Figure 1. The polymer electrolyte membrane serves as a medium for protons to be transported from the anode to the cathode.

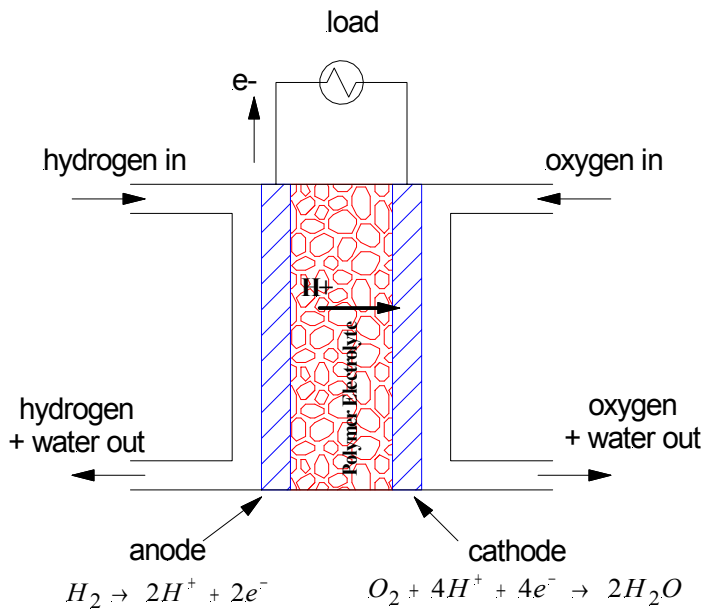

Figure 1. Hydrogen-oxygen PEM fuel cell. Hydrogen molecules dissociatively adsorb at the anode and are oxidized to protons. Electrons travel through an external load resistance. Protons diffuse through the PEM under an electrochemical gradient to the cathode. Oxygen molecules adsorb at the cathode, are reduced and react with the protons to produce water. The product water is absorbed into the PEM, or evaporates into the gas streams at the anode and cathode.

$$H_2 \rightarrow 2H^+ + 2e^- \qquad O_2 + 4H^+ + 4e^- \rightarrow 2H_2O$$

The proton conductivity is nearly independent of temperature when expressed as a function of water activity (water activity $a_w$={water pressure/vapor pressure of water}$P_w/P^o$=RH/100). The physical basis for proton conductivity can be explained as follows. At low water activity the PEM absorbs water that ionizes the sulfonic acid groups, which liberates protons that can hop between the fixed anions. The distance between ionized sites decreases with water uptake. Hopping frequency depends exponentially on distance so the proton conductivity increases exponentially with water activity. As the water activity increases and more water is taken up by the membrane the hydrophilic regions of the PEM swell gradually decreasing the resistance to proton transport in the membrane. Theoretical models have been fit to the proton conductivity due to membrane swelling. (water activity > 0.15).(Hsu and Gierke 1982; Eikerling, Kornyshev et al. 1997; Thampan, Malhotra et al. 2000; Paddison 2001)

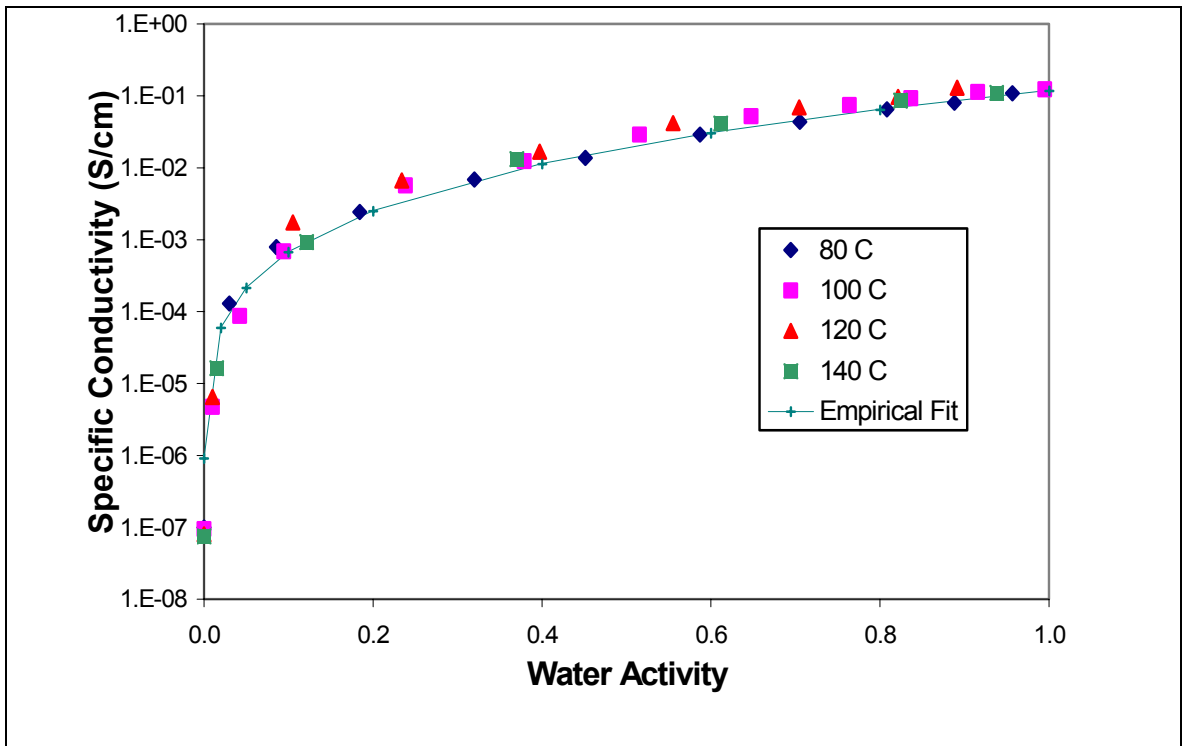

Figure 2. Conductivity of a Nafion 115 membrane as a function of relative humidity at temperatures ranging from 80 to 140°C. The empirical fit to the data ($\sigma=10^{-7}*\exp(14(RH/100)^{0.2}$ S/cm) was used to model fuel cell performance.(Yang 2003)

Membrane humidification is essential for a PEM fuel cell. If the membrane is not adequately humidified the proton conductivity will decrease, and the current output for the fuel cell will decrease proportionally. At the other extreme too much water will flood the electrodes, hindering gas transport to the membrane/electrode interface. The natural question from a chemical reactor viewpoint is what is the optimal level of water in the fuel cell, and how can we control that level. If humidification of the feed streams is required and how should the flow rates of the feed gases be adjusted to maintain adequate humidification in the polymer membrane? Additionally, how should the gas flow rates be adjusted as the load resistor on the fuel cell changes and alters the rate of water production?

The general approach to PEM fuel cell operation is to fix the humidity of the feed gases to the fuel cell near 100% relative humidity fixing the water content of the membrane. This approach treats water activity of the membrane as an adjustable parameter. However, water activity in a PEM fuel cell is a dynamic variable not an adjustable parameter. The fuel cell produces water, and the amount of water produced varies with the resistive load on the fuel cell. We treat the water activity as a system variable depending on the system parameters: the load resistance of the fuel cell, the temperature of the cell, the flow rates of the gases to the anode and cathode, and the humidity of the feed gases.

To elucidate the operation of a PEM fuel cell, we have constructed a simplified fuel cell reactor based on the concept of a stirred tank reactor borrowed from chemical engineering. The gas spaces above the anode and cathode, as shown in Figure 1, are designed for compositional uniformity, which permits the direct measurement of the reaction rate (current) as a function of the gas phase compositions at the anode and cathode.

We initiated our studies on the PEM fuel cell reactor kinetics with the simplest case possible: no humidification of the feed gases (autohumidification). The product water from the fuel cell reaction hydrates the membrane. We have operated PEM fuel cells continuously in the autohumidification mode for over 5000 hours (over 6 months!). We report here the conditions for steady state operation of a PEM autohumidification fuel cell, and demonstrate steady state multiplicity associated with ignition and extinction phenomena. We also report results where 5 steady states exist (three of them stable).

## II. EXPERIMENTAL

A schematic of the fuel cell for the kinetic studies in shown in Figure 3. The membrane-electrode-assembly (MEA) was pressed between two machined graphite plates and sealed with a silicon rubber gasket. Gas plenums of volume, V, of ~0.2 $cm^3$ were machined in graphite plates above a membrane area of ~1.5 $cm^2$. There were several pillars matched between the two plates to apply uniform pressure to the MEA. Hydrogen and oxygen were supplied from commercial cylinders (Airco) through mass flow controllers at flow rates, Q, of 1-10 sccm (mL/min). The residence times of the reactants in the gas plenums (V/Q) were greater than the characteristic diffusion time ($V^{2/3}/D$), assuring uniformity of the gas compositions. The cell temperature was controlled by placing the graphite plates between aluminum plates fitted with cartridge heaters connected to a temperature controller. The entire fuel cell assembly was mounted inside an aluminum box to maintain better temperature uniformity.

Pressure was maintained in the cell by placing spring loaded pressure relief valves (Swagelok) at the outlets. Tees were placed in the outlet lines (inside the aluminum box) with relative humidity sensors in the dead legs of the tees. The water content of the outlet streams was measured with humidity sensors (Honeywell HIH 3610), and the temperature at the humidity sensor was measured with a thermocouple in the gas line. The relative humidity sensors had to be sufficiently heated to avoid liquid condensation on the capacitive sensing element, but they also had to be kept below 85° to protect the amplifier circuit on the sensor chip.

The membrane-electrode-assembly (MEA) consisted of a Nafion™ 115 membrane pressed between 2 E-tek electrodes (these consist of a carbon cloth coated on one side with a Pt/C catalyst). The catalyst weight loading was 0.4 mg/$cm^2$. The electrodes were brushed with ~4 mg Nafion/$cm^2$ of solubilized Nafion solution before placing the membrane between them. The assembly was hot pressed at 130°C and 10 MPa. Copper foils were pressed against the graphite plates and copper wires were attached to connect to the external load resistor. The current and voltage across the load resistor were measured as the load resistance was varied. A 10-turn 0-20 Ω potentiometer was connected in series with a 10-turn 0-500 Ω potentiometer. The load resistance was varied from 0-20 Ω to obtain a polarization curve (IV). To examine the low current range of the polarization curve the resistance would be increased over the range of 0-500 Ω. The voltage across the load resistor was read directly by a DAQ board. The current through the load resistor was passed through a 0.2 ohm sensing resistor and the differential voltage across the sensing resistor was amplified by a factor of 100 with an Analog Devices AMP02 Instrumentation Amplifier and read by the DAQ board. An IV curve was typically collected and stored in ~ 100s.

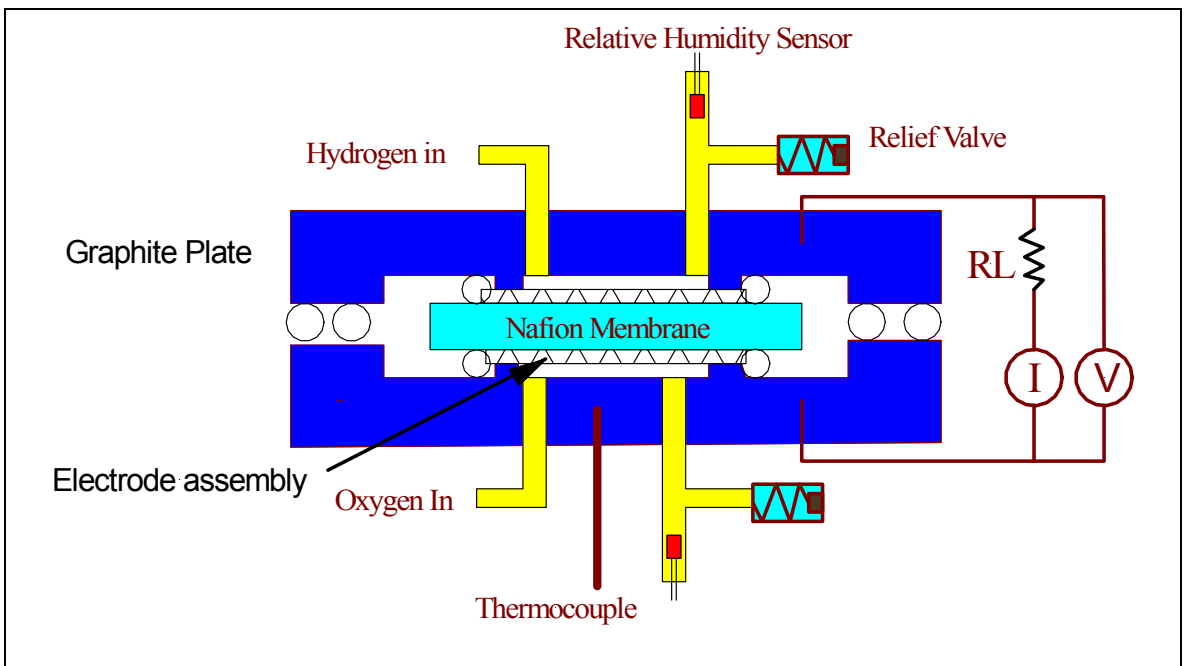

Figure 3. Schematic of the PEM fuel cell. The membrane-electrode-assembly is sandwiched between graphite plates with an open cavity for the gas above the electrodes. The graphite plates are placed between aluminum plates with cartridge heaters for temperature control.

Relative humidity measurements were taken intermittently for steady state conditions. Water condensation and/or temperatures above 80°C adversely affected the relative humidity sensors. A number of relative humidity sensors were destroyed at low gas flows when the water condensed on the sensors. Operating the cell at high pressures frequently resulted in water condensation on the humidity sensors. For these reasons we chose to limit our initial studies to atmospheric pressure. The temperature at the humidity sensors were below the cell temperature; at a cell temperature of 100°C the temperature at the sensors was ~65°C. We corrected the relative humidity measurements for gas flow rate, reactant conversion and temperature to determine the water removal rates.

## III. RESULTS

Three different types of measurements were taken in our studies:
    a. The steady state currents and voltages, and water removal in the anode and cathode streams were measured at variable load resistances and cell temperatures from 35-105°C with fixed flow rates of 10 mL/min of $O_2$ and $H_2$. These are time-consuming experiments as steady state required several hours to be established.
    b. Polarization data were obtained as functions of the flow rates of $O_2$ at the cathode and $H_2$ at the anode and cell temperature. In these experiments steady state would be obtained for a fixed load resistance and then the load resistance was varied over a short period of time with all other parameters fixed.
    c. Dynamic measurements of the cell current and voltage, and the relative humidity at the anode and cathode were followed during start-up of the fuel cell, and/or changing the load resistance.

These different experimental measurements illuminate the operation of the autohumidification PEM fuel cell as a chemical reactor. Care must be exercised in interpreting results from the fuel cell operation. Steady state is limited by the full equilibration of the membrane with the gas phase and the reaction rate (cell current). The membrane has a large capacity for water relative to the gas volume in the fuel cell. From our dynamic measurements we observed it took $10^3$-$10^4$ s for a cell to achieve steady state. Most polarization curves were obtained in $<10^2$ s corresponding to a fixed level of membrane hydration.

### III.1. *Fuel Cell Start-up and Ignition*

The start-up of an autohumidification PEM fuel cell was examined as a function of the initial water content of the Nafion membrane. The polymer membrane was dried by flowing dry oxygen through the cathode chamber at ~100 mL/min and dry nitrogen through the anode chamber at ~100 mL/min for ~12 hours. To humidify the membrane, the oxygen flow was shut off, and the nitrogen flow to the anode was passed through a water bubbler at room temperature. The relative humidity was measured at the outlet of the anode as a function of time to estimate the water uptake by the membrane. After hydrating the membrane to the desired level, the nitrogen flow was stopped. The fuel cell was heated to 50°C and the temperature was allowed to stabilize for ~ 1 hour. Hydrogen flow at 10 mL/min to the anode and oxygen flow at 10 mL/min to the cathode were established, and the current through the load resistor (set at 5 Ω) was measured as a function of time.

Figure 4 shows the current of the PEM fuel cell during startup. In Figure 4A the membrane was hydrated to different levels. The parameter λ is the number of water molecules taken up per sulfonic acid group in the Nafion membrane. The maximum water uptake has been measured in separate experiments and is ~12. (Springer, Zawodzinski et al. 1991; Thampan, Malhotra et al. 2000; Yang 2003) The results show classic ignition phenomena. For initial water loadings of <1.5 $H_2O/SO_3$ the fuel cell approaches an extinguished state, where the steady state current is very low (~0.1 mA).

When the initial water loading is >1.5 H$_2$O/SO$_3$ the fuel cell "ignites" and approaches a "high current" steady state current of ~ 125 mA. Ignition and extinction was also observed at other temperatures ranging from 35-105°C but we have not yet carried out a systematic investigation of the critical number of water molecules for ignition at those conditions.

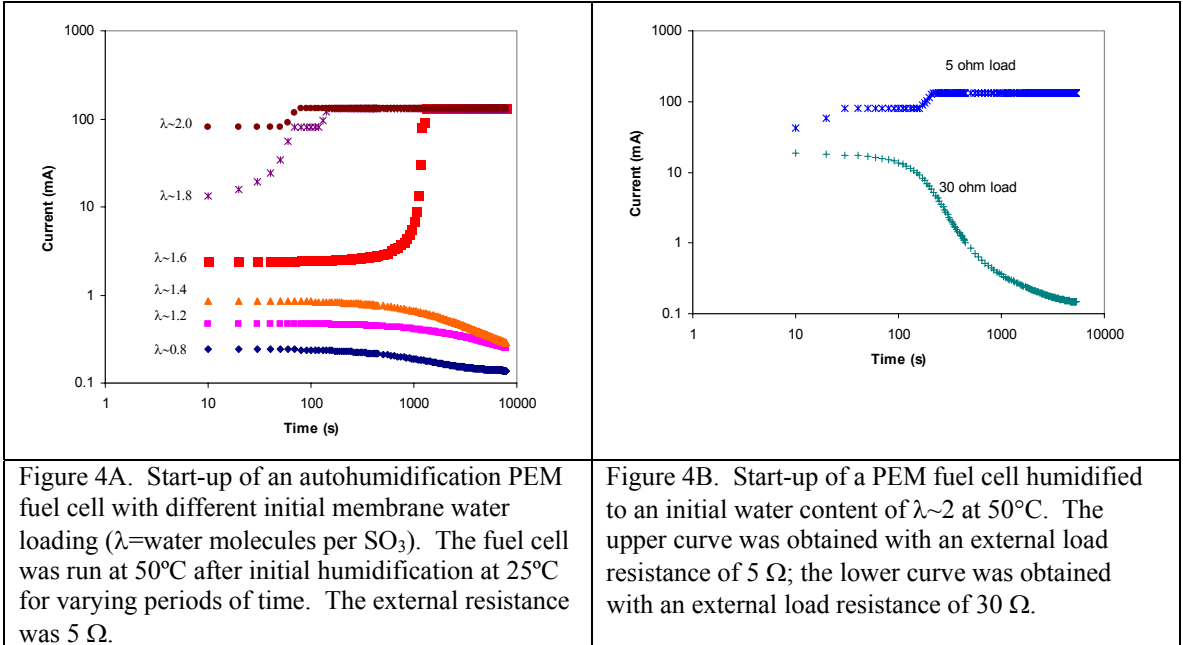

Figure 4A. Start-up of an autohumidification PEM fuel cell with different initial membrane water loading (λ=water molecules per SO$_3$). The fuel cell was run at 50ºC after initial humidification at 25ºC for varying periods of time. The external resistance was 5 Ω.

Figure 4B. Start-up of a PEM fuel cell humidified to an initial water content of λ~2 at 50°C. The upper curve was obtained with an external load resistance of 5 Ω; the lower curve was obtained with an external load resistance of 30 Ω.

Figure 4B shows the startup of the autohumidification fuel cell from the same hydration state with two different load resistances. At the higher load resistance the current is limited in the external circuit. Limiting the current reduces the production of water in the fuel cell. Water is removed by convection in the effluents from the anode and cathode. If the fuel cell is unable to produce enough water to sustain the hydration level in the membrane the fuel cell current is extinguished.

**III.2. *Steady State Operation of an Autohumidification PEM Fuel Cell***

Having established that a PEM fuel cell may be ignited and operated in the autohumidification mode, the steady state operation was examined as the temperature and load resistance were varied at fixed reactant flow. The hydrogen flow to the anode and oxygen flow to the cathode were both set to 10 mL/min (6770 nmol/s). The fuel cell temperature was set to the desired temperatures between 35-105°C, and the external resistance was set to 0.2 ohms. After allowing the fuel cell current to stabilize for several hours, the load resistance was set to 0.5 Ω and the fuel cell was allowed to stabilize for 1 hour. Normally the current would stabilize in ~1 minute, but we always waited for 1 hour to establish steady state. The relative humidity readings at the anode and cathode generally stabilized in less than 5 minutes. After recording the steady state current, and relative humidities the load resistance was increased by 2-3 Ω and the process was repeated. After increasing the load resistance to 15 Ω the fuel cell was allowed to stabilize an additional 12 hours to verify that the current and relative humidities were at steady state. A complete series of steady state measurements at a single temperature took a minimum of 3 days.

The relative humidity readings (RH) of the effluents from the anode and cathode were converted to mole fractions through global mass balances. The molar flow rates of the feed are $F^{in}$(mole/s), $i$ is the cell current (Amperes), $F$ is Faraday's constant (coulomb/mole), and $\xi$ is the fraction of water formed at the cathode that exits in the anode effluent.

$$\frac{RH_{anode}}{100} = x_{H_2O}^A = \frac{\xi \cdot i/2F}{F_A^{in} + \xi \cdot i/2F - i/2F}$$

$$\frac{RH_{cathode}}{100} = x_{H_2O}^C = \frac{(1-\xi) \cdot i/2F}{F_C^{in} + (1-\xi) \cdot i/2F - i/4F}$$

(1)

Equations (1) are solved for consistency of the value of $\xi$ for the relative humidity at the anode and the cathode. In all cases reported here the water was nearly equally partitioned between the anode effluent and the cathode effluent. Equipartitioning indicates that the water formed at the cathode equilibrates with the membrane and there is only a small gradient in water concentration across the membrane. The water removal rates at the anode and cathode are given by the molar flows of the effluents multiplied by the mole fraction of water in the effluent streams, and are given by equation 2.

$$F_{A,H2O}^{out} = \xi \cdot i/2F$$

$$F_{C,H2O}^{out} = (1-\xi) \cdot i/2F$$

(2)

Table I summarizes the steady state current, and water removal from both the anode and cathode as functions of the load resistance, and temperature. All these data are steady states – the fuel cell was operated in excess of 24 hours for the conditions tabulated to verify steady state operation. In particular the conditions at 95° and 105ºC were tested continuously for over 100 hours to verify the steady state operation. The results in Table I show that the PEM fuel cell was operated in an autohumidification mode over a wide temperature range (35-105ºC), with varying load resistances. All these steady states can be characterized as "ignited" except for operation at 105ºC with a 15 Ω load resistance, where the fuel cell current is extinguished.

**Table I**
**Auto-humidification PEM Fuel Cell Performance**

| Cell Temperature (°C) | Load Resistance (Ω) | Steady State Current (mA/cm²)* | Water Removed at Anode (nmol/cm²-s) | Water Removed at Cathode (nmol/cm²-s) |
|---|---|---|---|---|
| 35 | 0.5 | 164 (849) | 400 | 460 |
|  | 5 | 78 (404) | 205 | 200 |
|  | 15 | 31 (161) | 70 | 85 |
| 50 | 0.5 | 180 (932) | 430 | 520 |
|  | 5 | 80 (166) | 180 | 215 |
|  | 15 | 32 (166) | 90 | 85 |
| 65 | 0.5 | 176 (912) | 420 | 500 |
|  | 5 | 78 (404) | 175 | 220 |
|  | 15 | 31 (161) | 65 | 90 |
| 80 | 0.5 | 145 (751) | 350 | 400 |
|  | 5 | 76 (394) | 175 | 200 |
|  | 15 | 30 (156) | 80 | 80 |
| 95 | 0.5 | 130 (674) | 325 | 370 |
|  | 5 | 72 (373) | 160 | 200 |
|  | 15 | 30 (156) | 85 | 75 |
| 105 | 0.5 | 54 (280) | 125 | 160 |
|  | 5 | 31 (161) | 70 | 90 |
|  | 15 | <1 | <5 | <5 |

*values in parentheses is the water production rate in nmol/cm²-s

### III.3. *Fuel Cell Operation with Variable Reactant Flow Rates*.

The variables under operator control in the autohumidification PEM fuel cell are the load resistance, the fuel cell temperature, and the gas flow rates through the anode and cathode chambers. Steady state operation at fixed flow rates was reported above. Fuel cell operation as a function of reactant flow rates was characterized by the polarization curves at different temperatures. The polarization curve is the typical method of characterizing fuel cell performance. However, the polarization curves generally do not represent steady state performance of a PEM fuel cell. It takes approximately $10^3$ s for the membrane hydration to equilibrate with a change in the water production at the cathode. Unless the polarization curve was taken with small changes in the external load over the period of many hours the fuel cell will not be operating at steady state. If the polarization curve is taken in a period of ≤100 s the water content in the membrane will be approximately constant. The polarization curve under these conditions represents a constant level of membrane hydration, but this is not steady state.

Fuel cell operation with variable reactant flow rates was examined under conditions where the level of membrane hydration was kept approximately constant. After setting the fuel cell temperature, the hydrogen and oxygen flow rates were both set to 10 mL/min. The load resistance was adjusted to 0.5 ohm, and the fuel cell was run for 12 hours to establish steady state. The reactant flow rates were then adjusted to the desired conditions and the fuel cell operation was allowed to stabilize for 5 minutes. The load resistance was varied from 0.2-20 ohms in 100 s while the current through the load resistor and the voltage across the load resistor were measured. A typical polarization curve from the autohumidification fuel cell is shown in Figure 5A. The same data can be presented as a *Power Performance Curve*, where the power delivered to the load resistor (IV) is plotted as a function of the load resistance (V/I). Two key engineering variables of the PEM fuel cell are identified from the power performance curve, shown in Figure 5B. There is a clear maximum in the power from the fuel cell. The load resistance at the maximum power is equal to the effective internal resistance of the membrane-electrode-assembly at the operating conditions. (This equivalence is shown in the Appendix).

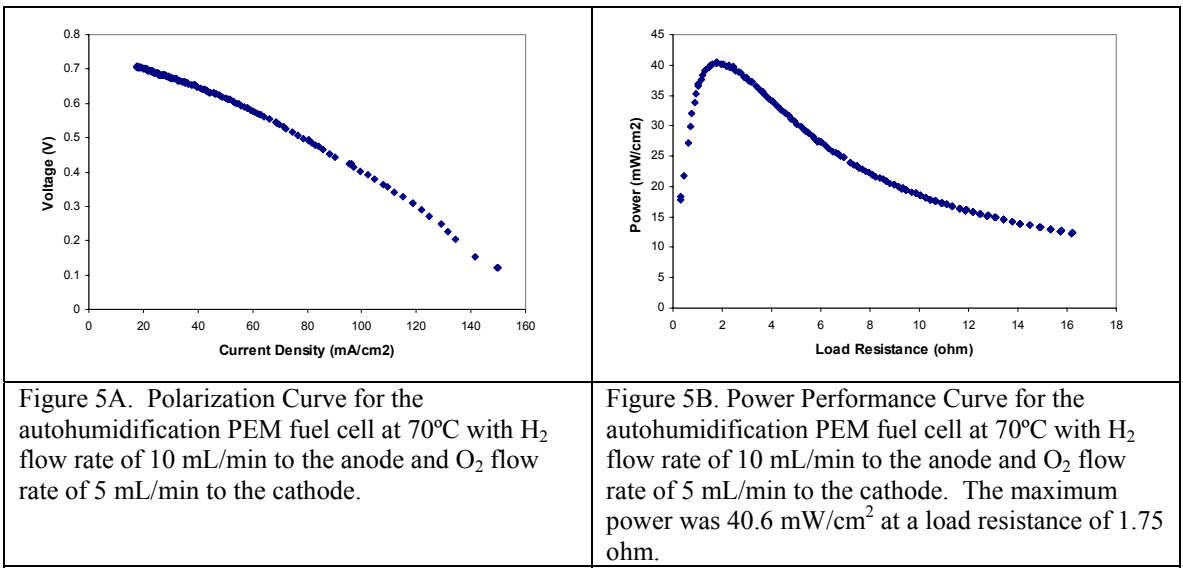

| Figure 5A. Polarization Curve for the autohumidification PEM fuel cell at 70ºC with $H_2$ flow rate of 10 mL/min to the anode and $O_2$ flow rate of 5 mL/min to the cathode. | Figure 5B. Power Performance Curve for the autohumidification PEM fuel cell at 70ºC with $H_2$ flow rate of 10 mL/min to the anode and $O_2$ flow rate of 5 mL/min to the cathode. The maximum power was 40.6 mW/cm² at a load resistance of 1.75 ohm. |
|---|---|

From the power performance curve we identify three figures of merit for comparing PEM fuel cell performance at different operating conditions,

     i.       fuel cell current at fixed load resistance
     ii.      maximum power at specified reactant flow rates and cell temperature
     iii.     load resistance at maximum power

Figure 6 shows the cell current as a function of the reactant gas flow rates for a load resistance of 5 Ω and cell temperature of 70ºC. At fixed oxygen flow the fuel cell current increased with hydrogen flow and plateaued. The cell current plateaued at ~120 mA for $H_2$ flow rates > 3 ml/min. The cell current was nearly independent of $H_2$ flow when the $H_2$ utilization was < 30%. Figure 6B shows that the fuel cell current was insensitive to the oxygen flow rate at fixed hydrogen flow for oxygen flow rates > 1 mL/min (this was also evident in Figure 6A where all three data sets at different $O_2$ flow rates fell on top of each other). These conditions correspond to oxygen utilization <50%.

We did not make quantitative measurements of the dynamic response of current to changes in flow rates, but we did make semiquantitative observations. The fuel cell current responded rapidly, within seconds, to changes in the hydrogen flow rate, but the current responded more sluggishly to changes in the oxygen flow rate. The current shut off within seconds of stopping the hydrogen flow, whereas it took more than a minute for the current to shut off when the oxygen flow was stopped. Experiments at 55ºC and 85ºC, were qualitatively identical to the results shown in Figure 6 for 70ºC. The current plateau achieved at different feed stoichiometries with a 5 Ω load resistance is documented in Table II. The maximum currents at 55ºC were approximately the same as at 70ºC, while the maximum currents were approximately 10 mA lower at 85ºC than at 70ºC. Because these measurements were not at steady state there is s

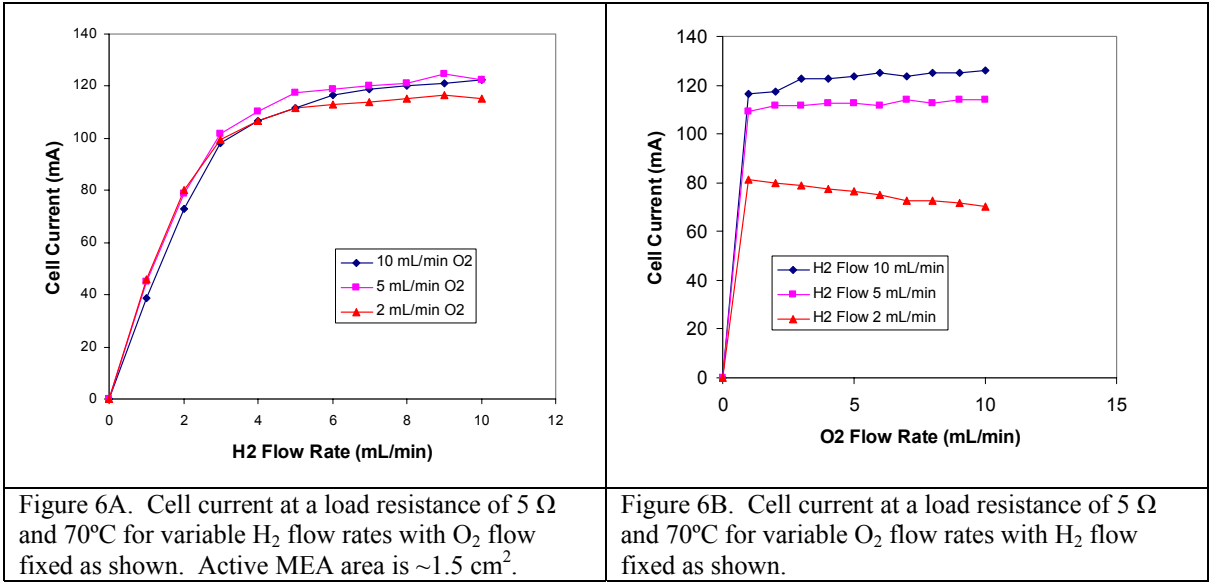

Figure 6A. Cell current at a load resistance of 5 Ω and 70ºC for variable $H_2$ flow rates with $O_2$ flow fixed as shown. Active MEA area is ~1.5 cm$^2$.

Figure 6B. Cell current at a load resistance of 5 Ω and 70ºC for variable $O_2$ flow rates with $H_2$ flow fixed as shown.

**Table II**
**Maximum Fuel Cell Currents in Autohumidification PEM Fuel Cell**

| Temperature (°C) | $H_2/O_2$ Feed Ratio | | | | |
| | 5.0 | 2.5 | 1 | 0.5 | 0.2 |
| | Cell Current (mA) | Cell Current (mA) | Cell Current (mA) | Cell Current (mA) | Cell Current (mA) |
| --- | --- | --- | --- | --- | --- |
| 55 | 118 | 119 | 122 | 120 | 100 |
| 70 | 115 | 122 | 122 | 114 | 82 |
| 85 | 100 | 114 | 110 | 100 | 70 |

We can get information about the hydration state of the membrane and the effects of flow rates on membrane dehydration from the Power Performance Curves. The Power Performance Curves give the maximum power and effective MEA resistance at the given feed flow rates and fuel cell temperature. Figures 7 and 8 plot the maximum cell power and the resistance at maximum power as functions of the flow rates at 85ºC. The results are very similar at 55ºC and 70ºC. The principal difference is that the maximum power output is slightly higher at the lower temperatures and the effective MEA resistances are slightly lower at the lower temperatures.

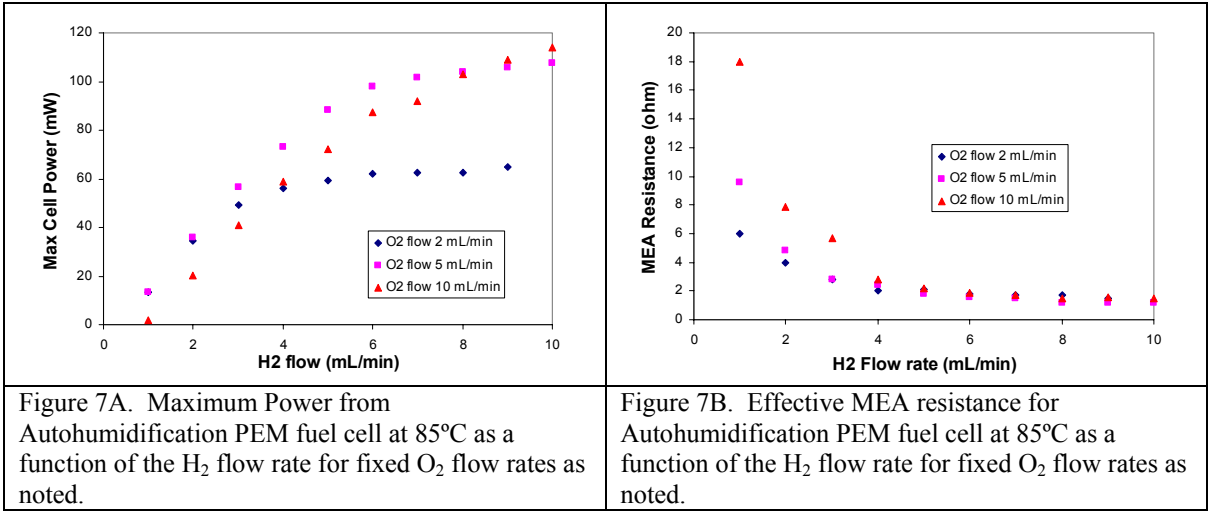

Figure 7A. Maximum Power from Autohumidification PEM fuel cell at 85ºC as a function of the $H_2$ flow rate for fixed $O_2$ flow rates as noted.

Figure 7B. Effective MEA resistance for Autohumidification PEM fuel cell at 85ºC as a function of the $H_2$ flow rate for fixed $O_2$ flow rates as noted.

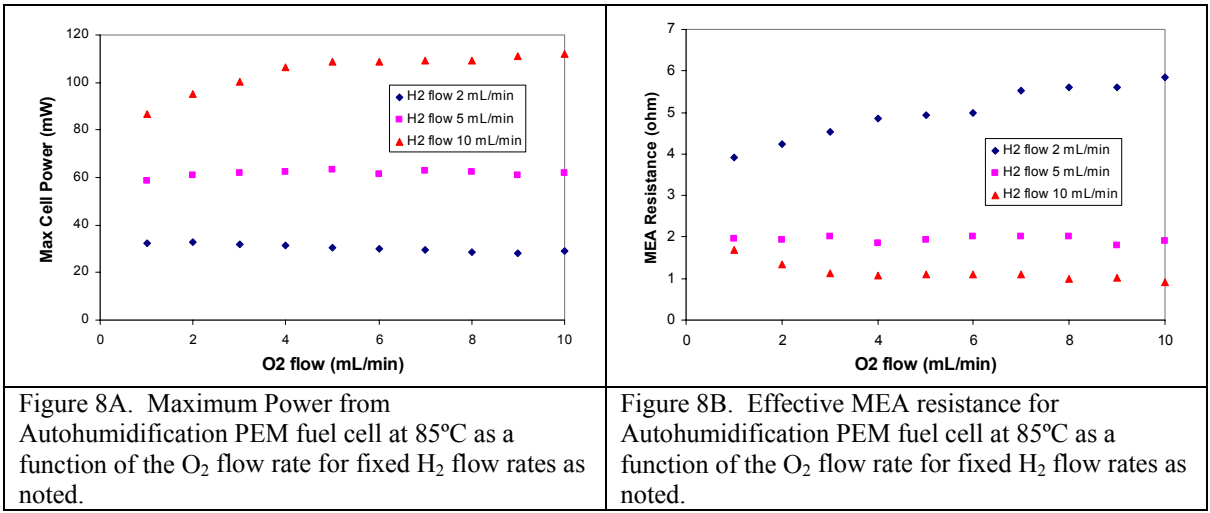

Figure 8A. Maximum Power from Autohumidification PEM fuel cell at 85ºC as a function of the $O_2$ flow rate for fixed $H_2$ flow rates as noted.

Figure 8B. Effective MEA resistance for Autohumidification PEM fuel cell at 85ºC as a function of the $O_2$ flow rate for fixed $H_2$ flow rates as noted.

Figure 7 shows two trends. When the oxygen flow is fixed, the maximum power saturates when the feed is fuel rich ($H_2/O_2>2$). Increasing the $H_2/O_2$ ratio beyond stoichiometric does not affect the maximum power, but increasing the total flow rate for the same stoichiometry resulted in increased power output. Second, the effective MEA resistance decreases with increased $H_2$ flow and plateaus when the feed is $H_2$ rich. Two effects combine to produce the highest power output. At high oxygen flow (substoichiometric feed), water is removed at the cathode faster, reducing the water content of the membrane. A low hydrogen flow limits the water production, again reducing the water content of the membrane. These effects are illustrated in Figure 7B where the MEA resistance increased dramatically at the lowest hydrogen flow rates and highest oxygen flow rates. At all the conditions represented in Figures 7 and 8 the utilization of the reactants at the maximum power production is always less than 40%, so the large increase in effective MEA resistance cannot be explained by depletion of the hydrogen.

The key trend shown in Figure 8 is that neither the power production nor the effective MEA resistance is very sensitive to the oxygen flow. The power increases with increased hydrogen flow, but is nearly constant with oxygen flow. At the lowest hydrogen flow rate the MEA resistance increases with $O_2$ flow probably the result of greater water loss from the cathode side of the membrane.

The results presented in this section provide guidance concerning the effects of reactant flow rates and temperatures on the water production rate in PEM fuel cells. However, unlike the results presented in the previous section, the cell currents and the membrane-electrode-assembly resistances are not steady-state measurements. We did not report any relative humidity measurements in this section because they were gradually changing during the course of the measurements of the polarization curves. The data presented in this section were obtained fast enough that the relative humidity at the anode or cathode did not change more than 5%, so the rate data correspond to "constant water content" in the membrane, fixed temperature and flow rates of the reactants.

### III.4. *Steady-state Multiplicity*

Results for ignition/extinction phenomena with the autohumidification PEM fuel cell were reported in section III.1. The ignited state could be extinguished by increasing the load resistance limiting the water production. It was expected that increasing the cell temperature would increase the rate of water removal and would also extinguish the fuel cell. As we report here in more detail, the behavior was not that simple.

The autohumidification PEM fuel cell was operated at steady state at 95ºC with a feed of 10 mL/min $H_2$ and 10 mL/min $O_2$ and a load resistance of 2.5 Ω. The steady state current was ~175 mA and was sustained over several days of operation. The current-voltage response of the fuel cell in a constant state of hydration was obtained in a period of 100 s. The polarization curve and Power Performance Curve at this state of membrane hydration are shown in Figure 9A and 9B respectively. We refer to this steady state as a high water content state as the effective MEA resistance is low corresponding to a high level of membrane hydration. The load resistance was increased to 20 Ω and the current was recorded as a function of time for a period of 8 hours (see Figure 10). The current dropped from 175 mA to 38 mA within seconds and then continued to decrease to a value of 35 mA after 300 s. The current showed a very slow decline of about 1 mA over the next 10,000 s, then showed a steeper decline of 3.5 mA over a period of 3000 s, and finally remained stable at a current of 31 mA for a period of 2 days. After 2 days a second set of current-voltage data was taken and the resulting polarization curve and Power Performance Curves are also shown in Figure 9. This final state is referred to as the low water content state; it has a higher MEA resistance corresponding to a lower level of membrane hydration.

The direction of the load resistance change was reversed. Steady state operation of the fuel cell operating was established with the load resistance set to 25 Ω. The current-voltage response of the fuel cell was recorded at a constant hydration level; the polarization curve and Power Performance Curve fell directly on top of the low water content curves previously recorded and shown in Figure 9.   The load resistance was reduced to 7 ohms and the current was recorded as a function of time (Figure 10B).  The current jumped immediately from 25 mA to 65 mA when the load resistance was decreased.  The current remained steady for almost 1500 s, and then jumped to 87 mA, where it remained steady for a period of more than 24 hours.  The current-voltage response of the fuel cell was recorded.  The polarization curve recorded after the new steady state was established fell almost directly on top of the polarization curve for the high water content shown in Figure 9A.

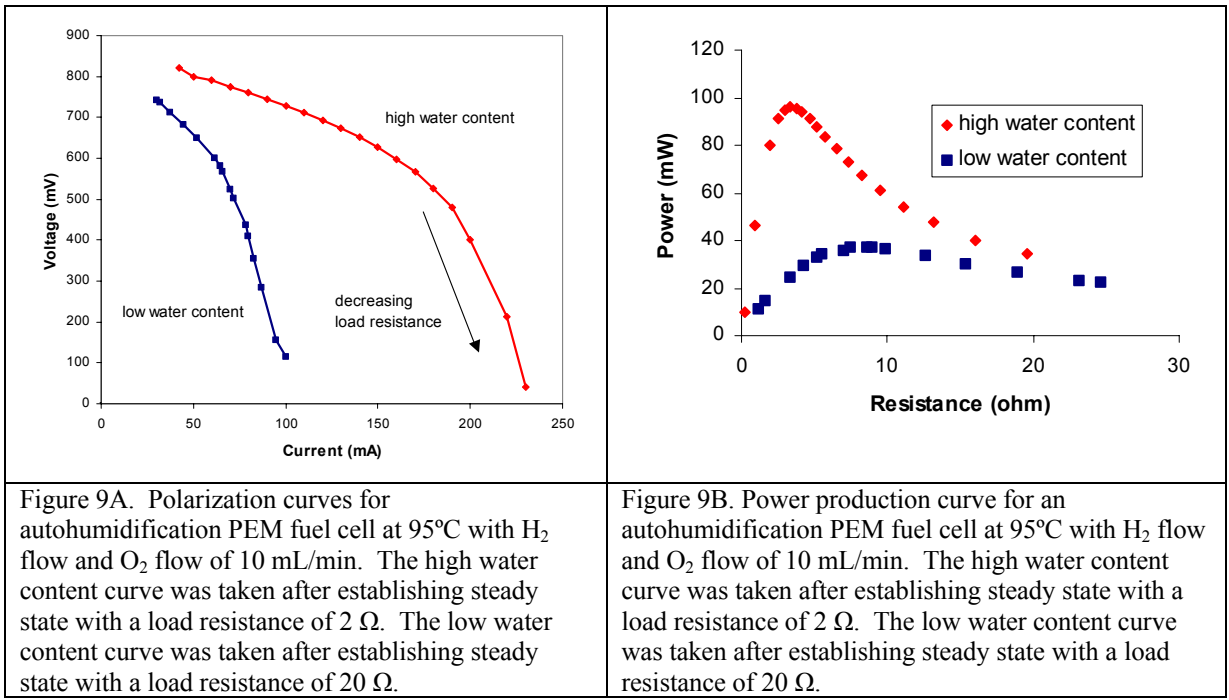

Figure 9A.  Polarization curves for autohumidification PEM fuel cell at 95ºC with $H_2$ flow and $O_2$ flow of 10 mL/min.  The high water content curve was taken after establishing steady state with a load resistance of 2 Ω.  The low water content curve was taken after establishing steady state with a load resistance of 20 Ω.

Figure 9B. Power production curve for an autohumidification PEM fuel cell at 95ºC with $H_2$ flow and $O_2$ flow of 10 mL/min.  The high water content curve was taken after establishing steady state with a load resistance of 2 Ω.  The low water content curve was taken after establishing steady state with a load resistance of 20 Ω.

These transitions between steady states were repeated five times over a period of two weeks and were highly reproducible.  The transitions between the two steady states were examined in more detail.  The relative humidity at the cathode and anode was monitored simultaneous with the current after changing the external load resistance.  Those results are shown in Figure 11.  The relative humidity at both the anode and cathode jumped at concurrently with the jumps of the fuel cell current.  When the current increased in response to decreasing the load resistance, the relative humidity at the cathode increased almost instantaneously, and the relative humidity at the anode increased with a time lag of ~100s.  When the cell current was constant the relative humidity at the anode and cathode remained constant. And when the current jump after 1500 s, the relative humidity at the anode and cathode jumped in tandem.  (The relative humidities shown in Figure 11 are the measured values at the sensors; these values have not been corrected for the temperature difference between the fuel cell and the sensor, the effluent flow rates, and the water partitioning between the anode and cathode. Those corrections are complex for transient behavior. Unfortunately, we did not have good measurements of the temperatures at the relative humidity sensors during these experiments so we do not attempt to report corrected values. However, the absolute values are not critical – the temporal correlation between relative humidity and fuel cell current is the essential feature we want to report here.)

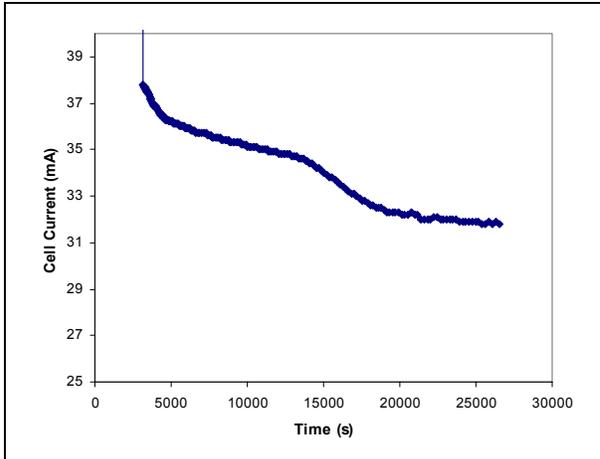

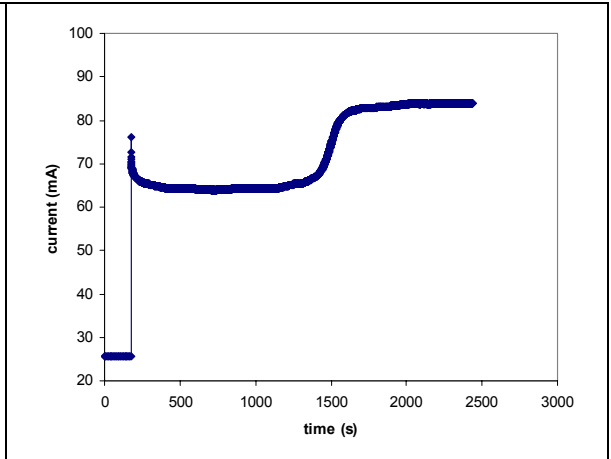

Figure 10A. Current response of PEM fuel cell at 95ºC to a step increase in the load resistance from 2 Ω to 20 Ω. The change was made at 3300 s. The initial current from 0 to 3300 s was 175 mA.

Figure 10B. Current response of PEM fuel cell at 95ºC to a step decrease in the external load resistance from 25 Ω to 7 Ω. The switch was made at 175 s.

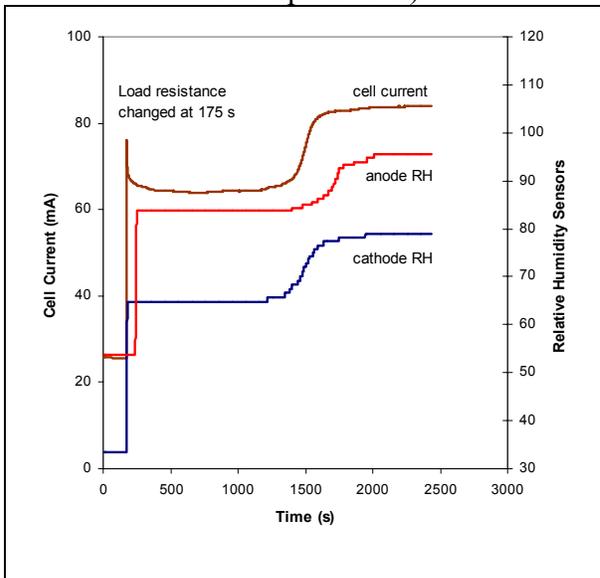

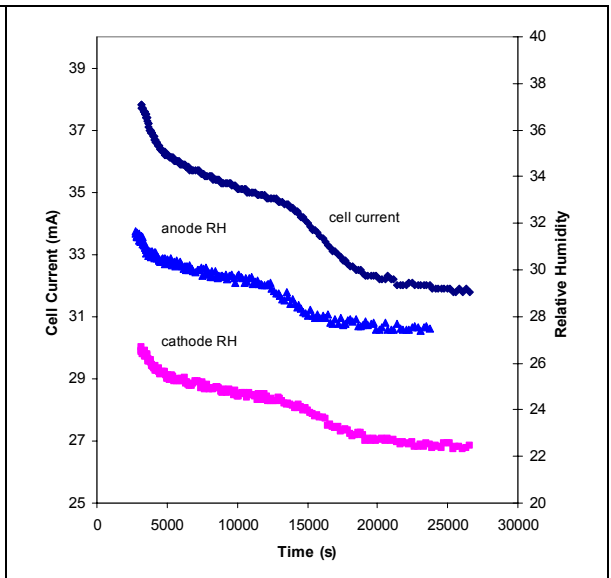

Figure 11 A. Fuel cell current at 95ºC and relative humidity in the anode and cathode effluents in response to a step decrease in the external load resistance from 25 Ω to 7 Ω. The switch was made at 175 s.

Figure 11B. Fuel cell current at 95ºC and relative humidity response to a step increase in the load resistance from 2 Ω to 20 Ω. The change was made at 3300 s. The current from 0 to 3300 s was 175 mA.

The relative humidity measurements also tracked the cell current when the experiment was switched from a low load resistance to a high load resistance. The single difference observed was the anode relative humidity appeared to decrease *prior* to the decrease in the cell current and the humidity at the cathode.

A series of measurements were taken examining the transitions between the "high water content" and "low water content" steady states. We started from a low resistance, with the fuel cell operating in the high water content steady state (so named because the MEA resistance is lower and the relative humidity at the anode and cathode are higher). The current-voltage response of the fuel cell was recorded. The polarization curve and the Power Performance Curve were plotted. The resistance was incrementally increased and the fuel cell current was allowed to stabilize over a period of 2 hours, after which another polarization curve was obtained. If the resistance of the MEA as determined from the maximum of the Power Performance Curve did not change by more than 0.2 ohm the load resistance was increased and the process was repeated. This process was used to find the load resistance at which a jump is observed in the MEA resistance. This procedure was employed to identify the critical load resistance where the fuel cell jumped from the high water content steady state to the low water content steady state.

This procedure was then reversed, starting from a high load resistance with the fuel cell operating in the low water content steady state the load resistance was decreased until the MEA resistance showed a large decrease indicative of the transition to the high water content steady state. The critical load resistances where the membrane transitions between a high and low humidification state are shown in Figure 12 for the autohumidification fuel cell. At load resistances between the two critical points the fuel cell could operate stably in either of two steady states, depending on how the steady state was approached. This is a classic hysteresis phenomenon associated with the existence of multiple stable steady states.

Measurements to identify the hysteresis loop were repeated at 50,65,80, and 95ºC. Two stable steady states were only observed at 80º and 95ºC; at 65º and 50ºC only one stable steady state was observed in our studies. The hysteresis loops at 80º and 95ºC are shown in Figure 12. The critical load resistance to jump from the high water content state to the low water content state decreases with increasing temperature; and the critical load resistance to jump from the low water content state to the high water content state increases with increasing temperature.

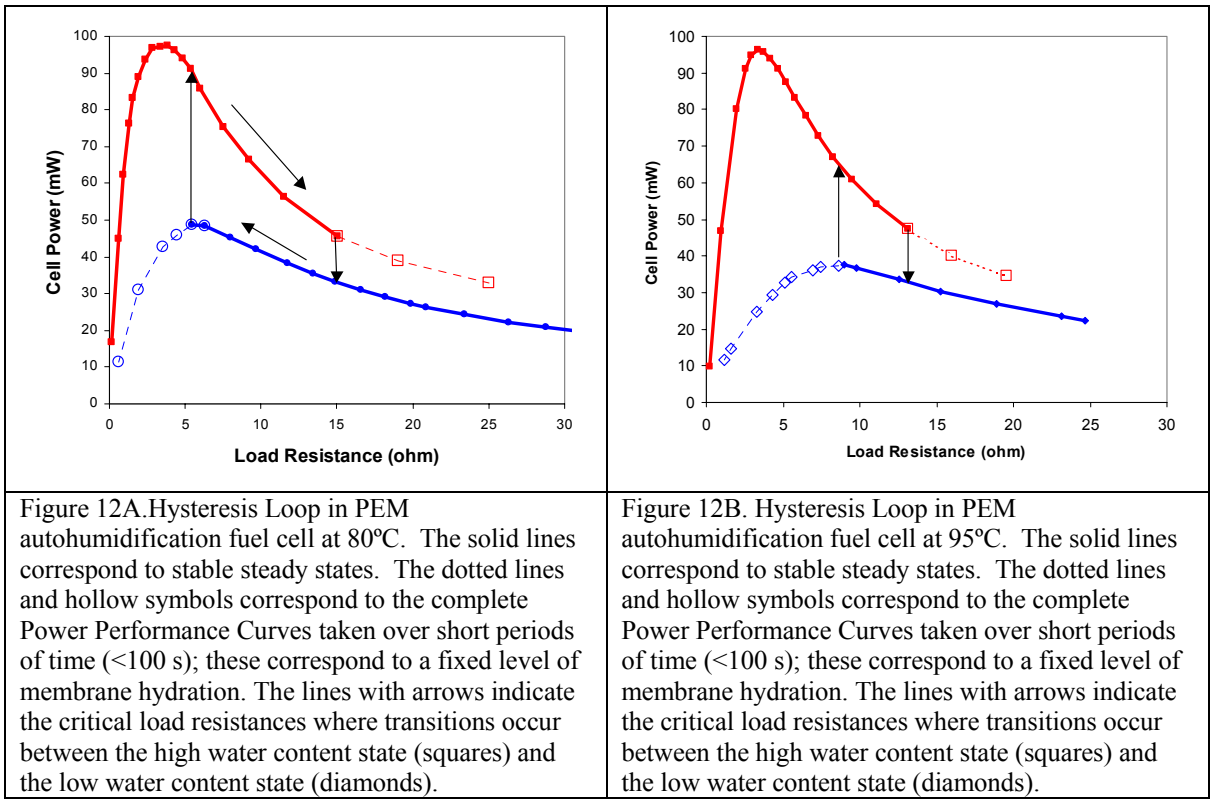

| Figure 12A. Hysteresis Loop in PEM autohumidification fuel cell at 80ºC. The solid lines correspond to stable steady states. The dotted lines and hollow symbols correspond to the complete Power Performance Curves taken over short periods of time (<100 s); these correspond to a fixed level of membrane hydration. The lines with arrows indicate the critical load resistances where transitions occur between the high water content state (squares) and the low water content state (diamonds). | Figure 12B. Hysteresis Loop in PEM autohumidification fuel cell at 95ºC. The solid lines correspond to stable steady states. The dotted lines and hollow symbols correspond to the complete Power Performance Curves taken over short periods of time (<100 s); these correspond to a fixed level of membrane hydration. The lines with arrows indicate the critical load resistances where transitions occur between the high water content state (squares) and the low water content state (diamonds). |
|---|---|

Several transient measurements were made of the "time to jump" between the high water content state and the low water content state as a function of the change in external load resistance (the time the current remained constant after the change in resistance shown in Figure 11A). These are tedious experiments and we have not completed a thorough study, but a few results are summarized in Table III. Two trends emerged: (i) the greater the difference between the critical load resistance and the applied load resistance the shorter the time to the jump from the low water content steady state to the high water content steady state; (ii) the transition time to jump from the low water content steady state to the high water content steady state is faster at lower fuel cell temperature.

Table III
**Transition Times to Jump for Low to High Steady States**

| Temperature | Starting load Resistance | Final load Resistance | Time to jump from low to high steady state (s) |
|---|---|---|---|
| 80 | 20 | 1 | 1100 |
| 80 | 20 | 5 | 2700 |
| 95 | 20 | 1 | 1200 |
| 95 | 20 | 5 | 3300 |

## IV. DISCUSSION

We have presented a novel experimental approach to PEM fuel cells, treating the fuel cell as a flow reactor. This approach uncovered ignition and multiple steady states in PEM fuel cells. The results also suggest that the transport processes in the membrane electrode assembly are not as well understood as previously thought. The novelty of the experimental approach here is that the fuel cell was not a scaled down version of a typical PEM fuel cell. Such fuel cells have long thin flow channels that traverse the membrane/electrode interface. The concentrations of the reactants and the water product all vary along the length of these channels, and the current measurements obtained from such a fuel cell are integrated values. The fuel cell employed in the experiments presented here operates as a well mixed stirred tank reactor. The gas phase compositions and membrane water content are all uniform, so the model fuel cell has well defined compositions. Although this is not the best configuration for optimal fuel cell performance, it is superior to extract the reactor kinetics, and understand the coupling of reaction, transport and physio-chemical properties that affect them.

### IV.1 Steady State Multiplicity in PEM Fuel Cells

The results presented here are contrary to some common assumptions about PEM fuel cells. It is frequently claimed that PEM fuel cells can only function with humidified feeds. Springer et al studied the operation of PEM fuel cells with humidified feeds where the humidification of the anode and cathode were varied.(Springer, Zawodzinski et al. 1991) Under their conditions they concluded that there was a net flux of water from the anode to the cathode due to electro-osmosis, and the PEM fuel cell could only function with a humidified hydrogen feed. This result is widely quoted and prominently stated in The Fuel Cell Handbook published by the Department of Energy.(EG&G Services 2000) Why were we able to operate a PEM fuel cell in an autohumidification mode?

We are not the first to report autohumidification PEM fuel cells. There are two previous reports we have identified for operation of PEM fuel cells without external humidification.(Watanabe, Uchida et al. 1996; Buchi and Srinivasan 1997) In addition there are PEM stacks that have incorporated internal humidification schemes.(H-Power began marketing a 500 W PEM stack in 2002 based on autohumidification operation, and United Technologies has a system for internal humidification). Our contribution is the identification of the critical parameters and mechanisms for humidity control in a PEM fuel cell that allow autohumidification.

We designed and operated our PEM fuel cell to have direct measurement of composition at the anode and cathode so composition could be directly correlated with the fuel cell current. A typical PEM fuel cell design has long thin serpentine flow channels at both the anode and cathode. The ratio of diffusive flow to convective flow in those channels, characterized by the dispersion number (diffusivity/((linear gas velocity)(channel size) $=D/ul$) is typically low ($\sim 10^{-3}$). This is typical of a plug flow tubular reactor (PFR); the gas composition varies along the length of the flow channels due to reaction with little axial dispersion due to diffusion. The measured fuel cell current is an averaged value of the local current over the length of the flow channels. In the PEM fuel cell employed here the dispersion number is ~1. This is typical of a stirred tank reactor (STR); the compositions at the anode and cathode are uniform due to diffusive mixing. The measured fuel cell current can be directly related to the gas phase composition at the anode and cathode. The STR fuel cell design employed here is ideal to characterize PEM fuel cell kinetics, and to examine cell dynamics. This is a one-dimensional system, which makes the analysis of the kinetics and dynamics much simpler. Results from the stirred tank reactor fuel call can easily be employed to model the typical flow channel systems in conventional PEM fuel cells. We don't advocate this as a method of designing fuel cells for maximum power density, but it does provide control in a fashion not possible in the standard PEM fuel cell design.

We also chose to operate the STR PEM reactor under conditions where the convective flows of the reactant gases at the anode and cathode were balanced with the fuel cell current. Fuel cell operation from low to high conversion of the reactants was examined by adjusting the external load resistance and the reactant flow. The advantages of the STR PEM fuel cell permitted us to explore details of PEM fuel cell operation that were previously unknown.

The STR PEM fuel cell permits direct probing of fuel cell operation with the level of hydration of the membrane. Many previous investigators have noted that with insufficient humidification of the PEM fuel cell the fuel cell did not function well and the maximum current derived from the fuel cell was very small. (Srinivasan, Manko et al. 1990; Blomen and Mugerwa 1993; Watanabe, Uchida et al. 1996; Buchi and Srinivasan 1997) There is a consensus that sufficient humidification of the feed was necessary for good operation, but we are aware of no study that has examined what level of humidification is necessary. To identify the necessary level of humidification we looked carefully at the initial water content in the membrane. Figure 4 clearly demonstrates critical water content was necessary to get the autohumidification PEM fuel cell to "ignite". The ignition/extinction phenomena can be easily understood from basic reaction engineering analysis.

Ignition and extinction result from balancing water production and water removal in the PEM fuel cell. Consider the mass balances on the hydrogen at the anode, oxygen at the cathode and water. It is assumed that there is no passage of either hydrogen or oxygen gases across the membrane.

Anode
$$\frac{Q_A^{in} P_H^{in} - Q_A^{out} P_H^{out}}{\mathsf{F}\,RT} = 0.5 i_{H^+} \qquad\qquad [3]$$

Cathode
$$\frac{Q_C^{in} P_O^{in} - Q_C^{out} P_O^{out}}{\mathsf{F}\,RT} = 0.25 i_{H^+} \qquad\qquad [4]$$

Water Balance
$$\frac{Q_A^{out} P_w^A + Q_C^{out} P_w^C}{\mathsf{F}\,RT} = 0.5 i_{H^+} \qquad\qquad [5]$$

The Q's represent molar flow rates through the anode and cathode chambers, P's represent the partial pressures in the chambers, and $i_{H+}$ is the proton current through the polymer membrane. We have assumed that no water is brought in with the hydrogen or oxygen feeds in the water balance. The current, or water production rate depends on the effective voltage across the membrane, V', and the resistance of the membrane-electrode-assembly for proton conduction. We approximate the effective cell voltage as the open circuit voltage less the activation potential, typically V'~0.85 V. The membrane-electrode-assembly resistance depends principally on the water content or water activity in the membrane and at the membrane/electrode interface. At high current densities or high reactant utilization the cell voltage and/or MEA resistance are also affected by mass transport resistances across the electrode and the ME interfacial resistance. Under most conditions we studied the mass transport resistances were negligible. Furthermore the membrane resistance changes exponentially with the water activity, which is a much larger effect than the changes in cell current due to mass transport resistances. We will neglect the mass transport resistances to the reaction rate and assume that Ohm's law gives the proton current, where the membrane resistance is a function of the water activity, $R_{membrane}(a_w)$, fitted to the experimental results shown in Figure 2.

Water Production Rate

$$0.5\, i_{H^+} = 0.5\, V'/(R_{membrane}(a_w) + R_{load}) \qquad [6]$$

The water generation, which is ½ the proton current, given by equation 6 is a sigmoidal function of water activity in the membrane due to the exponential decrease in resistance as a function of water content. A series of water production curves are traced out for different values of the load resistance in Figure 13. The membrane resistance at a given water content shows almost no temperature dependence (see Figure 2), so the water generation curves shown in Figure 13 may be considered as independent of temperature.

Steady state is established when the water removed by convection from the anode and cathode chambers is balanced by water production at the cathode. Water vapor transport from the membrane to the gas phases at the anode and cathode is an increasing function of the water activity in the membrane. For simplicity we will assume the water activity is uniform through the membrane, and the water vapor transport is given by an effective mass transfer coefficient that accounts for mass transport resistance across the membrane/electrode interface and mass transport through the porous electrode, $k_m$, multiplied by the water activity in the membrane, $a_w^{membrane}$, and the water vapor pressure at the fuel cell temperature, $P^o(T)$.

Water Removal Rate $\qquad\qquad F_w^{removal} = k_m a_w^{membrane} P^o(T) \qquad [7]$

The water removal curve corresponds to a straight line through the origin in Figure 13. The slope of that line increases with increasing temperature as the vapor pressure of water increases with temperature.

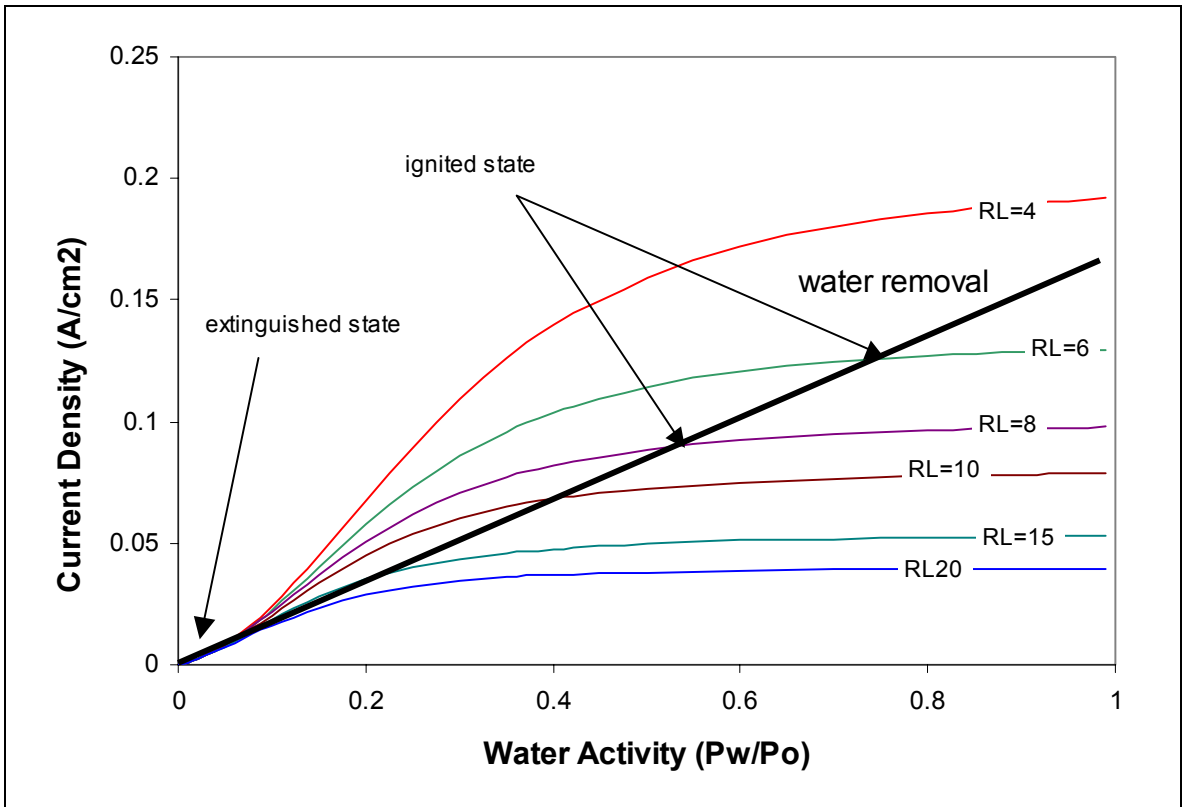

Figure 13. Water production (fuel cell current) and water removal rates for an auto-humidification PEM fuel cell. The water production is shown as a function of the external load resistance, and is defined by equation 6. The membrane resistance was an empirical fit of the data in Figure 2. The water removal line is given by equation 7, and the line shown corresponds to a cell temperature of 50ºC. The intersections of the water production curves and the water removal line represent steady states. The intersection near zero current density is an extinguished state, and the intersection at the higher current density of ~ 0.1 A/cm² is the ignited state. There is also an intersection of the water production and water removal curves around water activity of 0.08 and 0.015 A/cm². This is an unstable steady state.

Depending on the load resistance there exist either one or three intersections of the water production and water removal curves; these intersections correspond to steady states where water production is balanced by water removal. At high load resistance, R=20 Ω, there is only be a single low conversion steady state. At lower load resistances, R=6,8,10 Ω, there are three intersections of the water removed and water generated corresponding to three steady states. Increasing the cell temperature will increase the slope of the water removal line, and the range of load resistances for which three steady states occur will be shifted to lower resistances.

When three steady states exist only two of them are stable. At both the low and high current, fluctuations away from the steady state will result in the system returning to it. For example, at the high conversion a fluctuation increasing the water content in the membrane will result in the water removal being greater than the water generated, so the water content of the membrane will decrease returning the system to steady state. The middle steady state is unstable; positive fluctuations in the water content from that steady state will cause the system to generate even more water and move towards the high current steady state. Similarly, negative fluctuations will drive the system towards the low current steady state. This is precisely what was shown in Figure 4; when the membrane water content was greater than a critical valve the system to moved towards the high current steady state, and when the water content was below that critical value the system moved towards the low current steady state. There exists a locus of critical initial membrane water contents that are a function of load resistance and cell temperature that define a separatrix as to what state the fuel cell will evolve.

Steady state multiplicity displayed by the autohumidification PEM fuel cell (Figure 13) is analogous to that for an autothermal reactor with an exothermic reaction.(Uppal, Ray et al. 1974; Schmitz 1976) Steady state multiplicity in the autothermal reactor arises from positive feedback between a reaction product, heat, and the reaction rate. Heat from reaction raises the temperature, which facilitates further reaction. The feedback mechanism in the autohumidification PEM fuel is unique; the reaction product, water, facilitates a transport process (proton conduction), and only indirectly facilitates the reaction rate.

Figure 13 elucidates the ignition phenomena observed in Figure 4. The water removal line intersects the water production lines for the unstable steady state at $a_w$=0.08. When the initial water content of the membrane was below that in equilibrium with $a_w$=0.08 the water removal exceeds water production and the fuel cell extinguishes. At $a_w$=0.08 the water content in the membrane is ~1.5-1.6 water molecules/$SO_3$. (Yang 2003) The critical water loading for PEM ignition should be ~1.5-1.6 water molecules/$SO_3$, as observed. Figure 13 also reveals that above a critical load resistance of ~ 20 $\Omega$ the water production curve is always below the water removal curve; the fuel cell will never ignite if the load is too great. The results displayed in Figure 4 show the fuel cell was extinguished when the load resistance was increased to 30 $\Omega$.

Figure 13 provides a good qualitative account of ignition and multiple steady states in an autohumidification PEM fuel cell. A fully quantitative model will require a more detailed understanding of mass transfer resistances, electrode kinetics and membrane properties beyond our current understanding.

The model presented here is only valid for water activity < 1; it is limited to cases where only water vapor are present. Closer examination of Figure 13 suggests that at sufficiently low load resistance the ignited state would correspond to a water activity >1. This condition has been observed experimentally. When water production exceeds its removal by convection liquid water will condense in the fuel cell. Liquid water condenses in the pores of the electrode inhibiting mass transport of hydrogen and oxygen to the membrane/electrode interface and the current drops precipitously. After the flooding occurs the water removal by convection exceeds water production and eventually the gas flow removes enough water to permit the gases to more easily get to the membrane/electrode interface and the current increases. This creates a chaotic variation in the cell current over time. We observed such chaotic phenomena at a cell temperature of 35 C with a load resistance of 0.2 ohm, but have not yet characterized the details of the PEM fuel cell performance in this range of operating parameters.

**IV.2 Water Transport in PEM Fuel Cells**

Water partitioning from the effluent streams in the autohumidification PEM fuel cell was surprising. Based on past studies, such as those of Springer et al (Springer, Zawodzinski et al. 1991; Watanabe, Uchida et al. 1996; Buchi and Srinivasan 1997), we expected to see substantially more water in the cathode effluent than the anode effluent. Instead the water removal was nearly equal from both sides of the fuel cell. Water is made at the cathode/membrane interface. Water must diffuse through both the membrane and the porous anode to exit in the anode effluent. Water exiting from the cathode only diffuses through the porous cathode. The anode and cathode were identical in the PEM fuel cell employed here. The near equal partitioning of the water product indicates that the resistance to water diffusion in the electrodes must be greater than the resistance to water diffusion across the membrane.

**IV.3. Rate dependence on Reactant Concentrations**

There is a paucity of data relating the local gas composition at the anode and cathode to the fuel cell current. A number of studies have examined detailed electrode kinetics in well-defined systems. Our particular interest is to determine the dependence of the fuel cell current depends on the concentration of hydrogen at the anode and the concentration of oxygen at the cathode in the presence of the various mass transport resistances.

The results presented in Figure 6 showed there was little dependence of the fuel cell current on the flow rates of the reactants above as long as the utilization of the reactant was below 30% for hydrogen and 50% for oxygen. Figure 14 replots the data from Figure 6 but showing how the cell current depended on the partial pressure of the reactants at the anode and cathode. Figure 14A shows the cell current is nearly independent of oxygen pressure at the cathode over the range of 0.5-1 bar, when the hydrogen pressure at the anode is fixed. Both Figures 14A and 14B show the fuel cell current decreases with decreasing hydrogen pressure at the anode. The decrease in current is greater than anticipated based on thermodynamics, if all other parameters were fixed. These data are consistent with a kinetic limitation to reaction at the cathode, in agreement with other investigations.

The results presented here show that the fuel cell current is independent of oxygen pressure at the cathode over a wide range of oxygen pressures. The constancy of the reaction rate or current over such a broad range of reactant pressures simplifies the operation and control of the PEM fuel cell.

The PEM fuel cell operation does show substantial variation in the current with the hydrogen pressure at the anode. More data is needed to understand this better. The data suggest that protons and oxygen anions may be competing for adsorption sites on the cathode; high oxygen pressure at the cathode and low hydrogen pressure at the anode result in reduced proton activity and reduced reaction at the cathode. The data do indicate that at moderate to high utilization factors for hydrogen the fuel cell current falls below the thermodynamic predictions. In high area fuel cells with long flow channels accounting for the variation in reaction rates at the cathode along the length of the flow channel will be necessary.

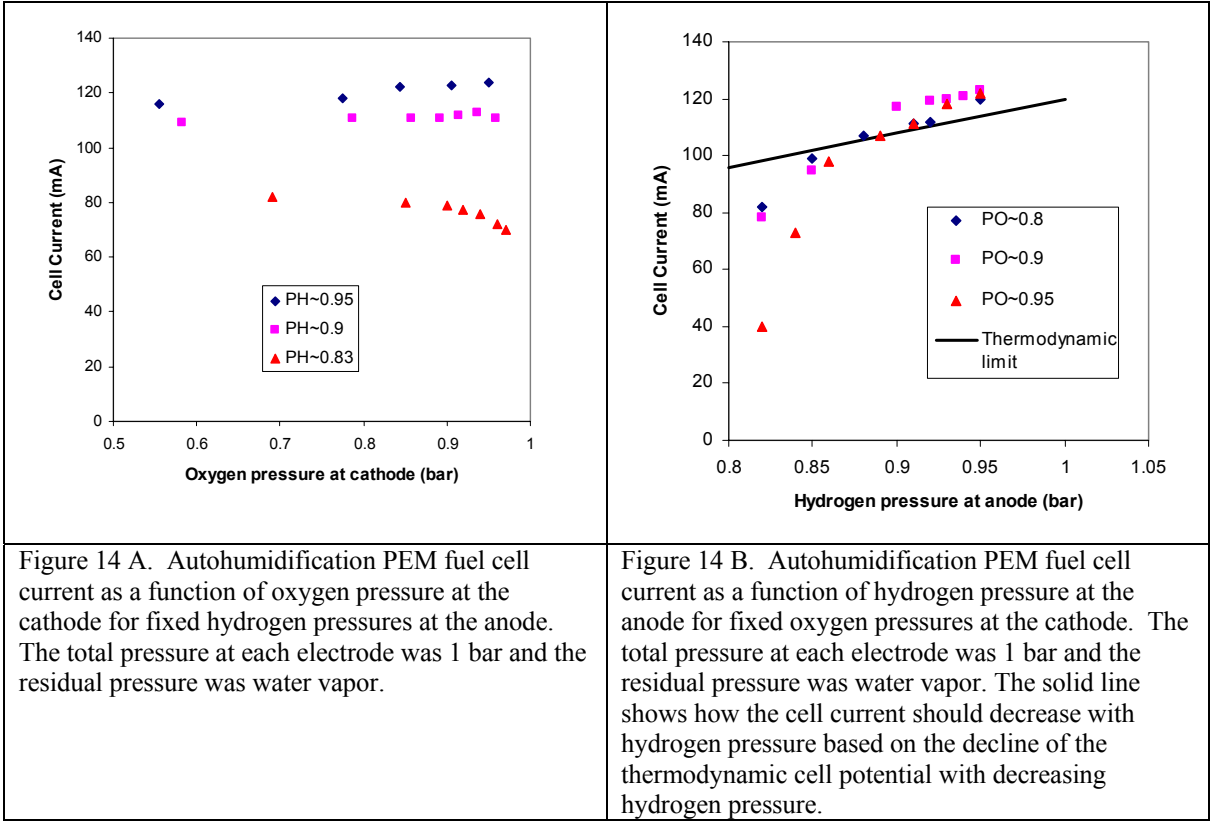

Figure 14 A. Autohumidification PEM fuel cell current as a function of oxygen pressure at the cathode for fixed hydrogen pressures at the anode. The total pressure at each electrode was 1 bar and the residual pressure was water vapor.

Figure 14 B. Autohumidification PEM fuel cell current as a function of hydrogen pressure at the anode for fixed oxygen pressures at the cathode. The total pressure at each electrode was 1 bar and the residual pressure was water vapor. The solid line shows how the cell current should decrease with hydrogen pressure based on the decline of the thermodynamic cell potential with decreasing hydrogen pressure.

## IV. 4. Existence of 5 Steady States

We initiated our work with the autohumidification PEM fuel cell expecting to observe three steady as illustrated in Figure 13. We were surprised to discover the existence of a third stable steady state. We have analyzed the experimental data to construct the water balance that accounts for five steady states. The water production curves show stepwise changes in the membrane resistance that has pointed to an explanation for the five steady states.

The steady state currents and membrane resistances as functions of load resistance and the hysteresis loops were used to construct the water production curve for conditions when three stable steady state exist. The procedure for construction of the water production curve at 95ºC is shown in Figure 15A. The steady state currents at different load resistances are shown as horizontal lines. For load resistances of $0.9 - 6.4 \ \Omega$ the stable steady state corresponded to the high membrane water content branch; these states designated by the solid horizontal lines at the upper right side of Figure 15A. At load resistances of $12.6 - 25.4 \ \Omega$ only the low membrane water content branch was stable; these are designated by solid horizontal lines at the bottom left of Figure 15A. When the load resistance was between 8 and 12 Ω either the high or low branch was stable, depending on the direction of approach. This middle region corresponds to where the hysteresis loop was observed; dashed horizontal lines designate those states.

There are two ignition points. The first ignition goes from the extinguished fuel cell operation to operation in the low water content branch. The second ignition point transitions between the low membrane water content state and the high membrane water content state. The first ignition point corresponds to the water activity in the membrane required for autohumidification; this was previously identified at a water activity of 0.08 (see Figure 13). The second ignition point can be bounded based on the water activity of the membrane of the low and high water content states. The membrane resistances of the high and low water content states can be obtained from the resistance at maximum power of the Power Performance Curve. The second ignition point must occur at a water activity greater than the water activity of the low water content state and less than the water activity of the high water content state. For a 10 Ω load resistance, the low water content state had an MEA resistance of 3.6 Ω ($\rho$=150 Ω-cm) and the high water content state had an MEA resistance of 1.2 Ω ($\rho$=50 Ω-cm). Based on the conductivity as a function of water activity in the membrane (Figure 2), this bounds the water activity of the second ignition point in the range of 0.25-0.35.

The experimentally determined MEA resistances of the low water content state and the high water content state were nearly constant (within 10%), independent of the load resistance. To fit the steady state current data as a function of load resistance water production curves must rise rapidly from zero to a horizontal segment corresponding to the low water content state, then a second rapid rise at the second ignition point followed by another horizontal segment corresponding to the high water content state. The water generation curves was empirically fit using equation 6 the water production, and expressing the membrane resistivity,$\rho$ (Ω-cm), with equation 8, and are shown in Figure 15B.

$$\rho = \begin{cases} 10^7 \exp(-14 a_w^{-0.2}) & a_w < 0.1 \\ 145 & 0.1 < a_w < 0.25 \\ 10^7 \exp(-14(a_w - 0.15)^{-0.2}) & 0.25 < a_w < 0.33 \\ 48 & 0.33 < a_w \end{cases}$$ [8]

The water removal curve is a straight line from the origin. It can be fixed by finding the intersection with the water production curves. For load resistances between 8-12 Ω the water removal line must intersect the water production curves at both the low and the high water content steady states. This is sufficient to fix the water removal line as shown in Figure 15B.

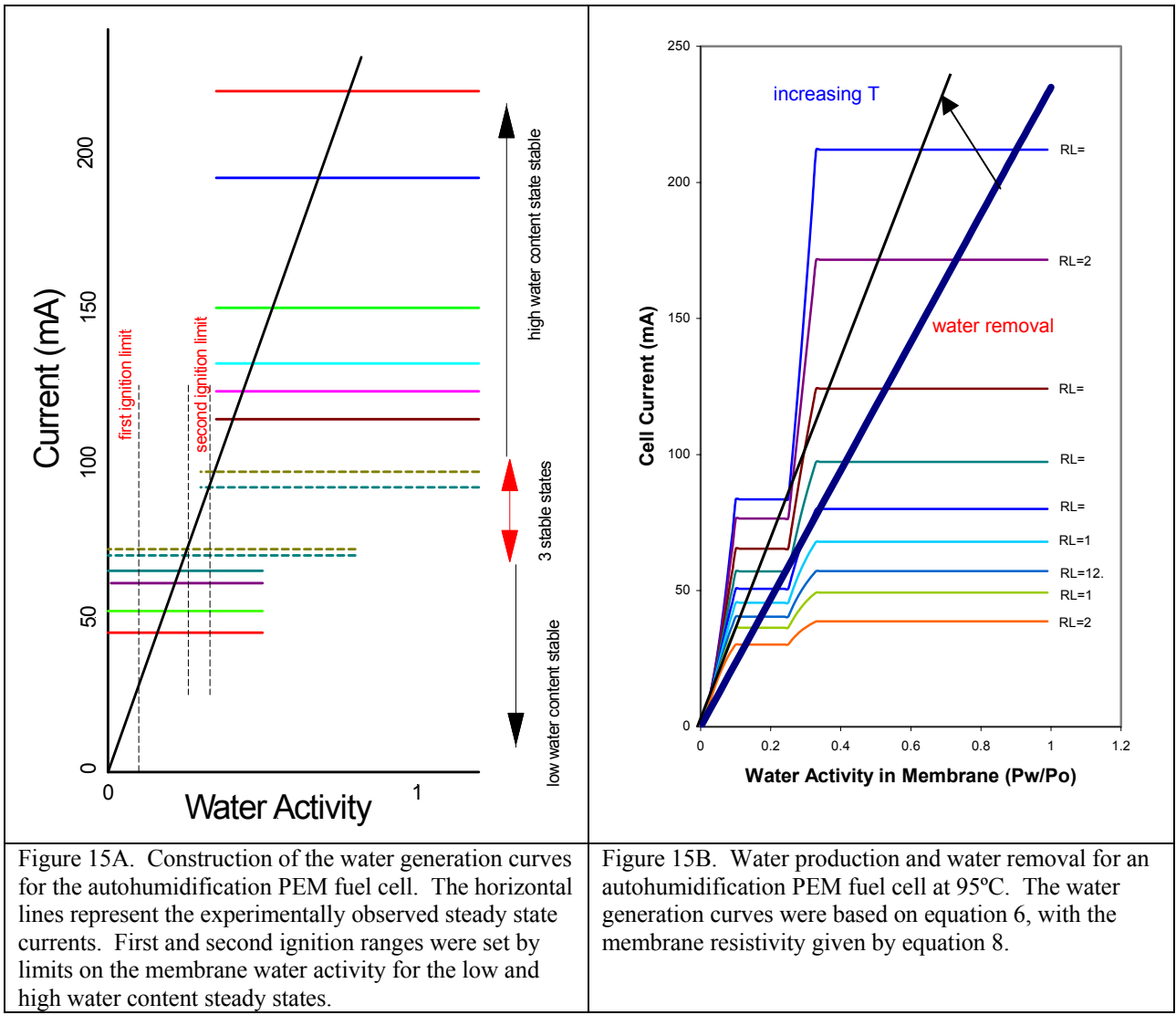

Figure 15A. Construction of the water generation curves for the autohumidification PEM fuel cell. The horizontal lines represent the experimentally observed steady state currents. First and second ignition ranges were set by limits on the membrane water activity for the low and high water content steady states.

Figure 15B. Water production and water removal for an autohumidification PEM fuel cell at 95ºC. The water generation curves were based on equation 6, with the membrane resistivity given by equation 8.

Figure 15B explains the origin of the hysteresis loops presented in Figure 12. Consider a fuel cell operating at steady state with a load resistance of 1 Ω; the fuel cell will be in the high water content steady state with a cell current of 220 mA. Increasing the load resistance to 2 ohms will result in a decrease in the cell current to 175 mA, but the cell continues to operate in the high water content state. As the load resistance is increased to 8 Ω the current will decrease smoothly to 80 mA. When the load resistance is increased to 10 Ω the high water content state is no longer stable. After increasing the resistance to 10 Ω the current will initially decrease to 68 mA, but eventually the fuel cell to drop to 52 mA where it will be in the low water content steady state. Increasing the load resistance further will cause the current to drop smoothly. If the load resistance is decreased from 10 Ω to 8 Ω the cell current will increase from 52 mA to 60 mA. The fuel cell will continue to operate on the low water content branch until the resistance is reduced to 6 Ω. When the resistance is decreased to 6 Ω the current will initially increase to 75 mA and eventually it will jump to 100 mA to its stable steady state. The data shown in Figures 10 and 11 correspond to the dynamics of these jumps.

We wish to stress the importance of distinguishing between true and "pseudo" steady states during testing of PEM fuel cells. The dynamics of transitions between steady states shown in Figures 10 and 11 show that it took $10^3 - 10^4$ s for a new steady state to be achieved after changing the external load resistance. PEM fuel cell characterization by current-voltage measurements is typically done in about $10^2$ s. During the time period typically allocated for a polarization curve measurement the water produced is not sufficient to equilibrate the water content of the membrane. A lower limit on the equilibration time for the membrane is given by the ratio of the water uptake per unit area of membrane divided by the water production at the cathode (this limit assumes all the water goes to membrane and none is lost to anode and cathode effluents). The water uptake per unit area is equal to some multiple of the number of sulfonic acid residues per unit area or membrane. The water production is ½ the current density.

$$\text{characteristic response time} = \frac{\text{sulfonic acid density}}{\text{current density/2}} \approx \frac{40\Delta\lambda}{j(A/cm^2)} \qquad [9]$$

The quantity $\Delta\lambda$ in equation 9 is the change in the number of water molecules per sulfonic acid group. For current densities $\sim$ 100 mA/cm$^2$ and a change in the hydration state of the sulfonic acids of 1 water molecules/SO$_3$ the characteristic time is 400 s. When we report polarization curves or Power Performance Curves these data are taken in <100 s, and represent an approximately constant hydration state of the membrane. To assure steady state operation of the fuel cell it is necessary to wait for several characteristic time periods, or several hours.

We were puzzled about the physics behind the five steady states and the dynamics of the transitions. In particular, the membrane resistance as a function of water activity given by equation 8 suggests there are some key physics that we did not recognize. The Power Performance Curves for the autohumidification PEM fuel cell operating in the high water content state as a function of temperature are shown in Figure 16A. The MEA resistance increases with temperature. Figure 16B shows the MEA resistance as a function of cell temperature for both the low water content state and the high water content state. The MEA resistances for the two branches diverge above 65ºC. Why should the five steady states have a start temperature, and what could be the physical cause for the phenomena?

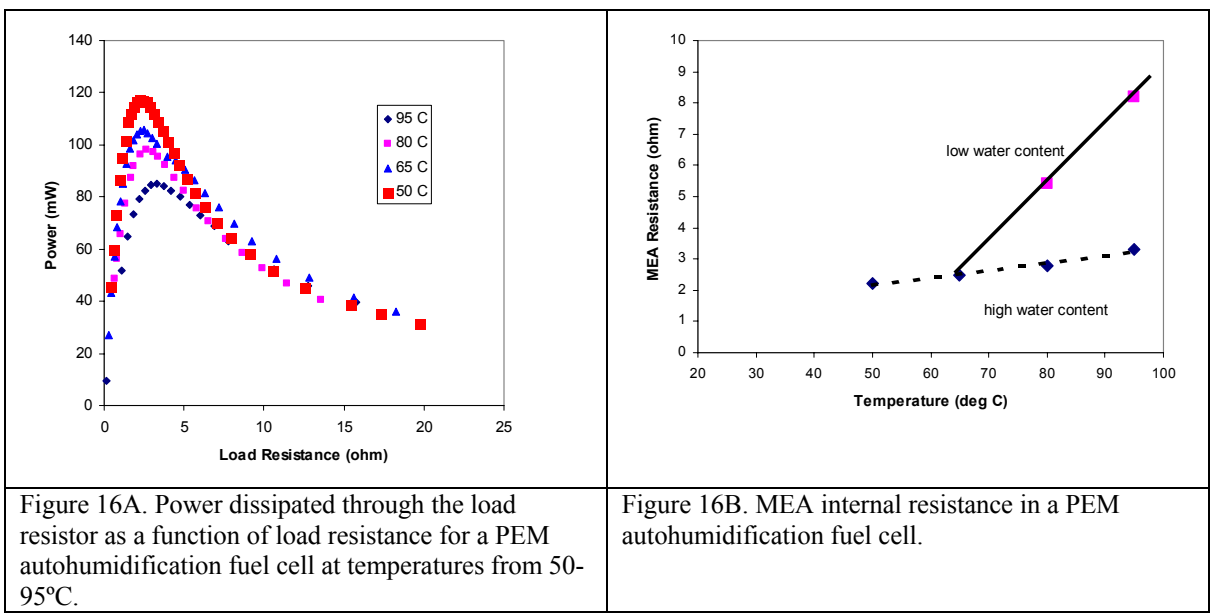

| | |
|---|---|
| Figure 16A. Power dissipated through the load resistor as a function of load resistance for a PEM autohumidification fuel cell at temperatures from 50-95ºC. | Figure 16B. MEA internal resistance in a PEM autohumidification fuel cell. |

We suggest the second ignition is due to a coupling of mechanical and chemical effects. The polymer membrane in the MEA is pressed between porous carbon electrodes, spatially confining the membrane. The entire MEA is pressed between graphite electrodes pressed against the MEA, as shown in Figure 17. As the membrane takes up water, the proton conductivity increases, but the membrane also swells. However, being spatially confined the membrane is not free to swell unlimited, it must do work to push the electrodes apart, or to swell into the porous electrode. Water uptake is hindered by the spatial constraints until the free energy associated with additional water uptake (swelling pressure) is sufficient to overcome the force to push the electrodes apart or penetrate into the porous electrodes. The energy required for swelling (or shrinking) of the membrane depends on the elastic properties of the membrane. Increasing temperature and increasing water content will both reduce the elastic modulus and glass transition temperature of the Nafion membrane. Above a critical temperature the swelling pressure of the membrane is reduced to the point where it is not sufficient to push the electrodes apart.

The effect of the spatially confined MEA on the water uptake is to pin the water activity at fixed values until the membrane swelling pressure overcomes the applied force of the electrodes. By pinning the water activity the proton conductivity is also pinned. This would result in a trend for the membrane resistance exactly as given in Equation 9.

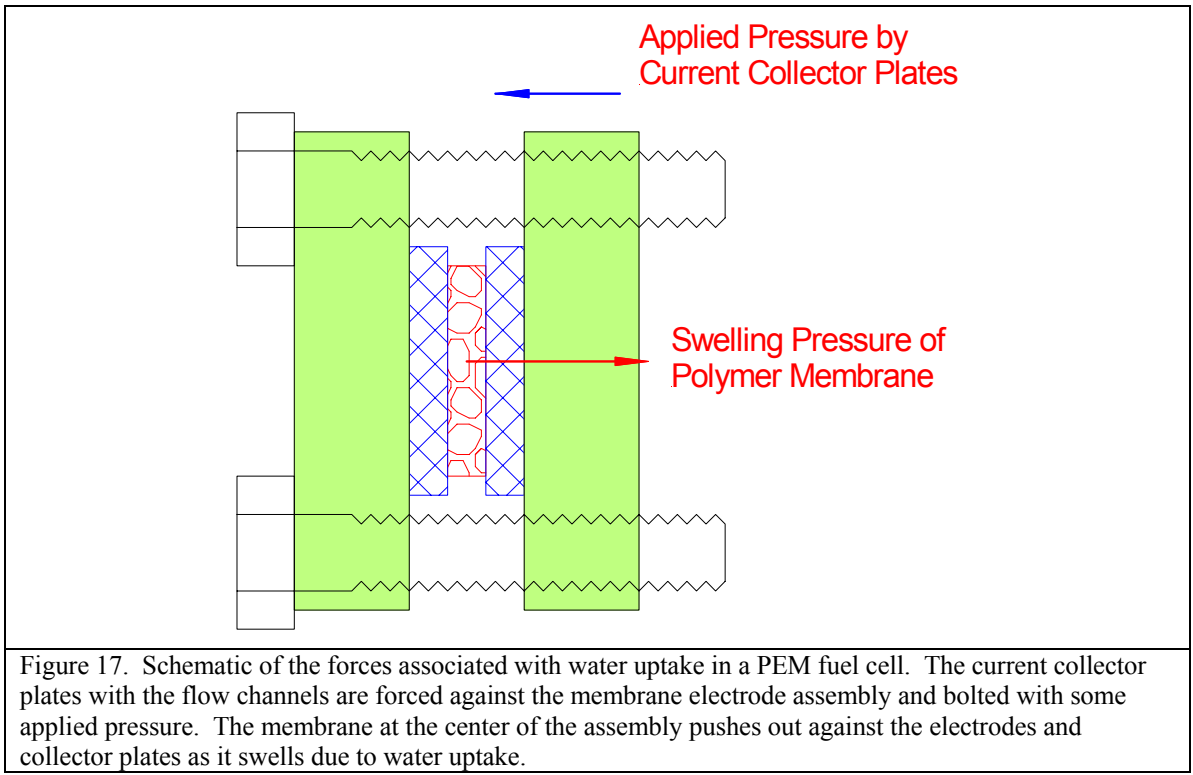

Figure 17.  Schematic of the forces associated with water uptake in a PEM fuel cell.  The current collector plates with the flow channels are forced against the membrane electrode assembly and bolted with some applied pressure.  The membrane at the center of the assembly pushes out against the electrodes and collector plates as it swells due to water uptake.

Proton conductivity in Nafion as a function of water activity has been studied, but these studies have always been done in unconstrained geometries.  The pressure loading of PEM electrodes may play a critical role in the performance of PEM fuel cells and this subject will be vital to proper design and control of PEM fuel cell stacks where both temperature and water activity will be varying.

## V. SUMMARY AND CONCLUSIONS

We have demonstrated the use of a model PEM fuel cell reactor that has permitted detailed studies of the kinetics, and the coupling of gas transport and electrochemical reactions.

(1) This reactor has permitted the unambiguous demonstration of an autohumidification PEM fuel cell between 30-105ºC.

(2)  The autohumidification PEM fuel cell displays the classic ignition/extinction phenomena associated with multiple steady states.  A critical initial water content in the membrane was required for the fuel cell to ignite.

(3) Feedback between the reaction product (water) and the transport of protons from the anode to the cathode was the cause of the steady state multiplicity.  This is a new phenomenon of coupling between a reaction product and a transport process leading to steady state multiplicity.

(4) The kinetics of water production were measured in the autohumidification PEM fuel cell.  For <50% utilization of hydrogen and <95% utilization of oxygen the fuel cell current was almost independent of the reactant flow rates.

(5) Above 65ºC the autohumidification PEM fuel cell displayed five steady states, (three of them stable) with hysteresis between a high water content state and a low water content state as the load resistance was varied. The five steady states were attributed to the coupling between chemical and mechanical properties of the polymer membrane due to swelling.

## References

Blomen, L. J. M. J. and M. N. Mugerwa, Eds. (1993). Fuel Cell Systems. New York, Plenum.

Buchi, F. N. and S. Srinivasan (1997). "Operating proton exchange membrane fuel cells without external humidification of the reactant gases - Fundamental aspects." Journal of the Electrochemical Society 144(8): 2767-2772.

EG&G Services, P., Inc. (2000). Fuel Cell Handbook. Morgantown, WV, US Department of Energy: 312.

Eikerling, M., A. A. Kornyshev, et al. (1997). "Electrophysical properties of polymer electrolyte membranes: A random network model." Journal of Physical Chemistry B 101: 10807-10820.

Hsu, W. Y. and T. D. Gierke (1982). "Ion Clustering and Transport in Nafion Perfluorinated Membranes." Journal of the Electrochemical Society 129(3): C121-C121.

Paddison, S. J. (2001). "The modeling of molecular structure and ion transport in sulfonic acid based ionomer membranes." Journal of New Materials for Electrochemical Systems 4(4): 197-207.

Schmitz, R. A. (1976). "Multiplicity, Stability, and Sensitivity of States in Chemically Reacting Systems - A Review." Advances in Chemistry 148: 156-211.

Springer, T. E., T. A. Zawodzinski, et al. (1991). "Polymer electrolyte fuel-cell model." Journal of the Electrochemical Society 138(8): 2334-2342.

Srinivasan, S., D. J. Manko, et al. (1990). "Recent advances in solid polymer electrolyte fuel cell technology with low platinum loading electrodes." Jounal of Power Sources 29: 367-387.

Thampan, T., S. Malhotra, et al. (2000). "Modeling of conductive transport in proton-exchange membranes for fuel cells." Journal of the Electrochemical Society 147(9): 3242-3250.

Uppal, A., W. H. Ray, et al. (1974). "Dynamic Behavior of Continuous Stirred Tank Reactors." Chemical Engineering Science 29(4): 967-985.

Watanabe, M., H. Uchida, et al. (1996). "Self-humidifying polymer electrolyte membranes for fuel cells." Journal of the Electrochemical Society 143(12): 3847-3852.

Yang, C. R. (2003). Performance of Nafion/Zirconium Phosphate Composite Membranes in PEM Fuel Cells. Deparment of Mechanical Engineering. Princeton NJ, Princeton University.
.

**Appendix**

**Power Performance Curve**

Fuel cell performance has typically been evaluated by the polarization curve where the voltage (V) between the electrodes is measured as a function of the current (I) through the external circuit. To measure the polarization curve, the external load is varied resulting in a change in the current passing through the fuel cell. A typical polarization curve is shown in Figure A1. The information contained in the polarization curve can be presented in different ways. We have found that for characterizing and evaluating the performance of a PEM fuel cell the power dissipated through the external load (IV) as a function of the external load ($R_L$=V/I) provides a helpful alternative presentation of the data. The power performance curve is shown in Figure A2.

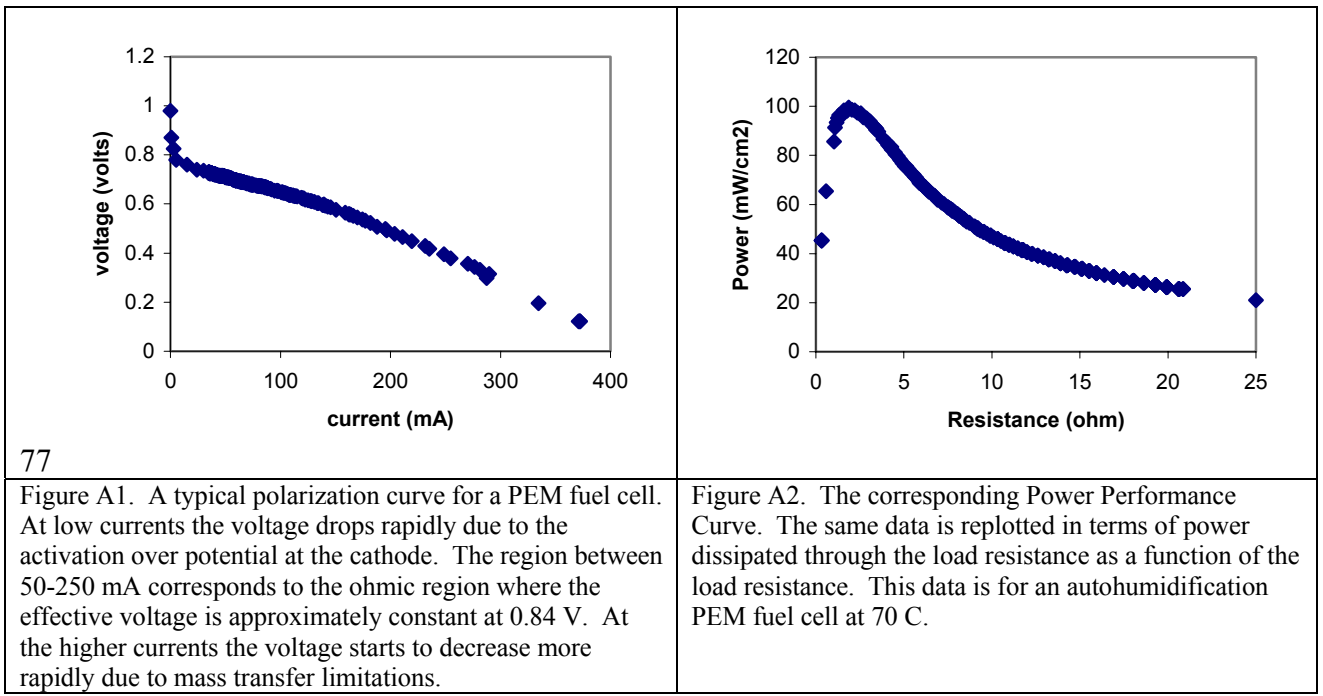

77

| Figure A1. A typical polarization curve for a PEM fuel cell. At low currents the voltage drops rapidly due to the activation over potential at the cathode. The region between 50-250 mA corresponds to the ohmic region where the effective voltage is approximately constant at 0.84 V. At the higher currents the voltage starts to decrease more rapidly due to mass transfer limitations. | Figure A2. The corresponding Power Performance Curve. The same data is replotted in terms of power dissipated through the load resistance as a function of the load resistance. This data is for an autohumidification PEM fuel cell at 70 C. |
|---|---|

Considering an equivalent circuit of a fuel cell, shown in Figure A3, reveals the utility of the Power Performance Curve representation. The fuel cell has been represented as a simple battery and resistance in series. During use, the fuel cell is where the voltage is approximately constant, equal to the open circuit voltage less the junction potentials and the over potential, V'=$V_{oc}$-$\eta$. This effective voltage is ~0.85 V, and will vary slightly with the MEA fabrication, choice of electrocatalysts etc.

When operating the fuel cell in the ohmic region the power dissipated in the load resistance is given by equation A1.

$$P = IV = \frac{(V')^2 R_L}{(R_m + R_L)^2} \qquad\qquad\qquad [A1]$$

The power has a maximum value when the external load resistance matches the membrane resistance, $P_{max}=(V')^2/4R_L$ when $R_L=R_m$. For engineering purposes one wants to run a fuel cell at high power density, so it is desirable to run near the maximum power.

Power Performance Curves are particularly useful in characterizing the hydration level of the membranes in PEM fuel cells. It is easy to identify the membrane resistance corresponding to maximum power, which in turn can be used to find the water activity in the membrane.

A key feature of the PEM fuel cell employed here was its small size (area ~1 cm2); the membrane resistance was ~ 1 Ω, and the cell currents were ~ 200 mA. Under these conditions a simple potentiometer could serve as the external resistance to record the power performance curve. For larger area PEM fuel cells the currents are larger and the resistances are significantly smaller, so that more complex devices are required to manipulate the resistive load.

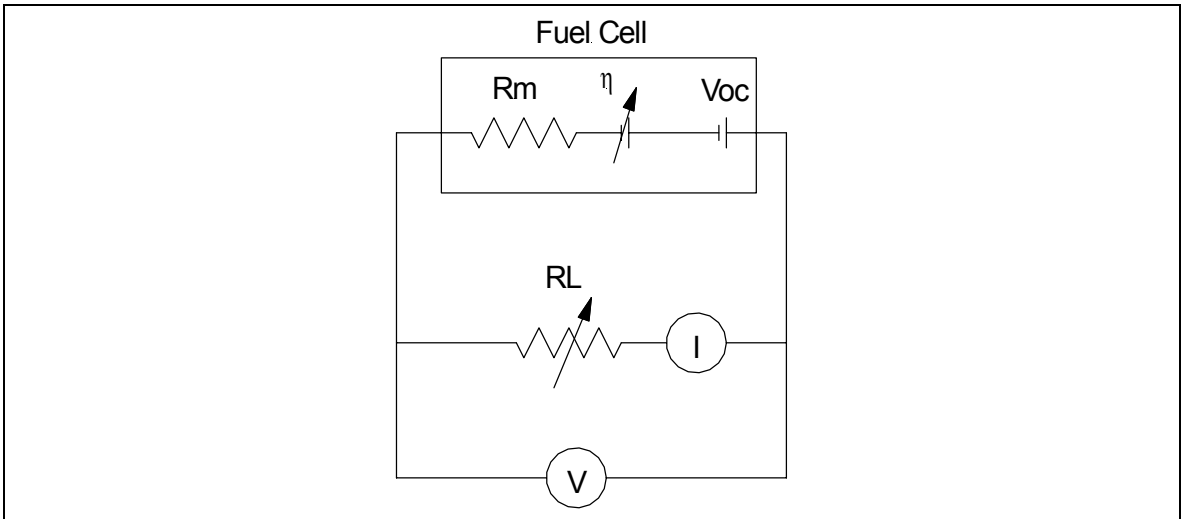

Figure A3. Equivalent circuit for PEM fuel cell characterization. Membrane resistance, Rm, a battery with open circuit voltage Voc and a variable potential representing the junction potentials and the overpotential associated with the cathode, replaces the fuel cell. The battery drives a current through a variable external load resistance $R_L$. The current through the load resistance (I) and the voltage across the load resistance (V) are measured as the load resistance is varied.